\documentclass[fleqn,usenatbib]{mnras}

\usepackage{newtxtext,newtxmath}

\usepackage[T1]{fontenc}
\usepackage{ae,aecompl}

\usepackage{graphicx}	
\usepackage{amsmath}	
\usepackage{cite}
\usepackage{textgreek}

\title[ALMA observations of Fomalhaut\,C]{ALMA imaging of the M-dwarf Fomalhaut\,C's debris disc}

\author[P. F. Cronin-Coltsmann et al.]{
\parbox{\textwidth}{Patrick F. Cronin-Coltsmann,$^{1,2}$\thanks{E-mail: patrick.cronin-coltsmann@warwick.ac.uk }
Grant M. Kennedy,$^{1,2}$
Paul Kalas,$^{3,4,5}$\\
Julien Milli,$^{6,7}$
Cathie J. Clarke,$^{8}$
Gaspard Duch\^ene,$^{3,7}$
Jane Greaves,$^{9}$\\
Samantha M. Lawler,$^{10}$
Jean-Fran\c cois Lestrade,$^{11}$
Brenda C. Matthews,$^{12,13}$\\
Andrew Shannon,$^{14}$
Mark C. Wyatt$^{8}$}
\\
\\
\parbox{\textwidth}{
$^{1}$Department of Physics, University of Warwick, Gibbet Hill Road, Coventry, CV4 7AL, UK\\
$^{2}$Centre for Exoplanets and Habitability, University of Warwick, Gibbet Hill Road, Coventry CV4 7AL, UK\\
$^{3}$Department of Astronomy, University of California, Berkeley, CA 94720, USA\\
$^{4}$SETI Institute, Carl Sagan Center, 189 Bernardo Ave., Mountain View CA 94043, USA\\
$^{5}$Institute of Astrophysics, FORTH, GR-71110 Heraklion, Greece\\
$^{6}$European Southern Observatory (ESO), Alonso de C\'ordova 3107, Vitacura, Casilla 19001, Santiago, Chile\\
$^{7}$Univ. Grenoble Alpes, CNRS, IPAG, F-38000 Grenoble, France\\
$^{8}$Institute of Astronomy, University of Cambridge, Madingley Road, Cambridge CB3 OHA, UK\\
$^{9}$School of Physics and Astronomy, Cardiff University, 4 The Parade, Cardiff CF24 3AA, UK\\
$^{10}$Campion College, University of Regina, Regina, SK S4S 0A2, Canada\\
$^{11}$LERMA, Observatoire de Paris, PSL Research University, CNRS, Sorbonne Universit\'es, UPMC Univ. Paris 06, 75014 Paris, France\\
$^{12}$University of Victoria, 3800 Finnerty Road, Victoria, BC V8P 5C2, Canada\\
$^{13}$National Research Council of Canada Herzberg, 5071 West Saanich Road, Victoria, BC V9E 2E7, Canada\\
$^{14}$LESIA, Observatoire de Paris, Universite PSL, CNRS, Sorbonne Universit\'e, Universit\'e de Paris, 5 place Jules Janssen, 92195 Meudon, France\\
}}

\date{Accepted XXX. Received YYY; in original form ZZZ}

\pubyear{2020}

\begin{document}
\label{firstpage}
\pagerange{\pageref{firstpage}--\pageref{lastpage}}
\maketitle

\begin{abstract}
Fomalhaut\,C (LP\,876-10) is a low mass M4V star in the intriguing Fomalhaut triple system and, like Fomalhaut\,A, possesses a debris disc. It is one of very few nearby M-dwarfs known to host a debris disc and of these has by far the lowest stellar mass. We present new resolved observations of the debris disc around Fomalhaut\,C with the Atacama Large Millimetre Array which allow us to model its properties and investigate the system's unique history. The ring has a radius of 26\,au and a narrow full width at half maximum of at most 4.2\,au. We find a 3$\sigma$ upper limit on the eccentricity of 0.14, neither confirming nor ruling out previous dynamic interactions with Fomalhaut\,A that could have affected Fomalhaut\,C's disc. We detect no $^{12}$CO J=3-2 emission in the system and do not detect the disc in scattered light with HST/STIS or VLT/SPHERE. We find the original Herschel detection to be consistent with our ALMA model's radial size. We place the disc in the context of the wider debris disc population and find that its radius is as expected from previous disc radius-host luminosity trends. Higher signal-to-noise observations of the system would be required to further constrain the disc properties and provide further insight to the history of the Fomalhaut triple system as a whole.

\end{abstract}

\begin{keywords}
binaries: general -- circumstellar matter -- planetary systems -- stars: individual: LP 876-10 -- stars: individual: Fomalhaut -- submillimetre: planetary systems
\end{keywords}



\section{Introduction}

The Fomalhaut system, one of the brightest in the night sky, has been subject to much observation, simulation and theoretical hypothesising over the past 35 years. A wide triple system, it comprises A4V star Fomalhaut\,A as well as K4V TW PsA (Fomalhaut\,B) at a 57,400 au separation, and M4V LP 876-10 (Fomalhaut\,C) at a 158,000 au separation \citep{Mamajek13}. The system is just 7.7 pc distant and 440 Myr old. The historic interest in the system can be attributed to two factors that are not necessarily unrelated. 

Firstly, both Fomalhaut\,A and C possess detectable debris discs. That is, we detect the presence of gas poor dust rings around the host stars. This dust is inferred to be continually produced by a collisionally evolving parent planetesimal population and not leftover from the protoplanetary disc \citep[e.g.][]{Hughes18,Wyatt08}.
The disc around Fomalhaut\,C was initially detected with Herschel PACS \citep{Kennedy13}; it was not spatially resolved but a temperature of 24K and radius of $\sim$20-40 au were estimated. Not much more about the disc could be discerned until it was recently observed with ALMA, as this paper will discuss. However, the Fomalhaut\,A debris disc has been clearly resolved in scattered light with HST \citep{kalas2005} as well as in the far-infrared with Herschel \citep{Acke2012} and in the millimetre with ALMA \citep{macgregor17,Boley12,White17}. These observations identify the belt as a sharply defined ring at a radius of $\sim$135 au, the centre of which is offset from the location of Fomalhaut\,A. The sharp definition of the edges and offset together imply a highly apsidally aligned population of planetesimals with a coherent eccentricity of $0.12 \pm 0.01$. Such disc morphologies are typically interpreted as the result of the action of a perturbing planet \citep{Wyatt99}. At first this perturbing planet seemed to be the directly imaged exoplanet candidate Fomalhaut\,Ab, a point-like object identified in HST observations \citep{Kalas08}. However, the point source could not be detected in the infrared and possessed a stellar-like colour, suggesting the flux originates from scattered stellar light and casting doubt on the hypothetical planet's nature \citep{Currie12}. Further HST observations proved that the object was on a highly eccentric orbit that is incapable of sculpting the disc into its present morphology \citep{Kalas13}. \citet{Kalas08} propose the point source is a low mass planet with a large circumplanetary ring system. A planet with a collisional swarm of irregular satellites has also been proposed and discussed \citep{Kennedy11,Tamayo14,Kenyon14}. But it has also been hypothesised that the point source is just a transient dust cloud \citep{Janson12,Kenyon14,Tamayo14,Lawler15,Gaspar20}.

Thus a separate planet must be invoked to drive the eccentricity of Fomalhaut\,A's debris ring for a planet driven scenario, however to date a second planet has not been identified in the system despite several searches \citep{Kenworthy13,Currie13}. \citet{Quillen06} predict this belt-shaping planet to have a mass of 0.04 -- 0.14 M\textsubscript{Jup} and \citet{Chiang09} predict a planet mass of 0.5 M\textsubscript{Jup}. On the additional assumption that Fomalhaut\,Ab was scattered into its current orbit by this putative planet, \citet{faramaz15} constrain a belt-shaping Fomalhaut\,Ac mass to 0.25 -- 0.5 M\textsubscript{Jup}.

Alternatively, simulations \citep{Lyra13} have shown gas-dust interaction could also organise dust into tight, eccentric rings. This can occur through instabilities within the disc \citep{Klahr05,Besla07} but requires a significant gas presence. Herschel PACS observations failed to detect C II and O I emission lines that would have been detected had the necessary quantities of gas been present in Fomalhaut\,A's disc \citep{cataldi15}. \citet{matra17} do detect the presence of CO in Fomalhaut\,A's disc using ALMA, but not in sufficient amounts to generate the necessary instabilities.

Past stellar interactions provide another mechanism for the generation of disc eccentricities, be this a flyby from an external star or the action of companion stars within the system. The action of flybys has long been investigated both in general theory \citep[e.g.][]{Kenyon02, Jilkova16} and in application to specific interesting systems, such as HD 141569 \citep[e.g.][]{Ardila05, Reche09} and HD 106906 \citep[e.g.][]{Rodet17, DeRosa19, Rodet19}.

In addition to the eccentric belt around Fomalhaut\,A, the system's unique orbital configuration provides a second point of interest. The wide orbits of Fomalhaut's stellar companions constitute sufficient angular momentum to preclude a common protostellar core fragmentation scenario. The system cannot have unfolded as per the model of \citet{Reipurth12}, as an angular momentum exchange resulting in a third star moving to a distant orbit requires the tightening of an inner binary. Stellar capture during the original cluster dispersal resulting in two wide companions is a viable history, but relies on two independently low probability events both occurring. The current wide separations also call into question the degree to which the system is bound and how it has evolved over its 440 Myr lifetime. The magnitude of the orbital period and the relatively meagre orbital velocities have prevented any definitive knowledge of the precise orbital configuration and trajectories from being surmised, yet several dynamical models for the system have been posited.

This paper considers whether new observations of the debris disc around Fomalhaut\,C with ALMA can provide evidence that Fomalhaut\,A's own eccentric planetesimal belt and the triple system's large stellar separations are connected through the system's dynamical history. Namely, our hypothesis is that if Fomalhaut\,A's eccentricity is due to previous interactions with Fomalhaut\,C, then Fomalhaut\,C's belt may be similarly affected and also show an eccentricity.
This paper presents and discusses previous works on the dynamics of the Fomalhaut system and further motivations for ALMA observations of Fomalhaut\,C in \S\ref{sec:motfac}, followed by a description of those observations in \S\ref{sec:obsv}. We then present an analysis of the observations in \S\ref{sec:ResAn} and discuss implications for our understanding of the system as a whole as well as the wider context of M\,star debris discs in \S\ref{sec:discussion}.

\section{Dynamical Hypotheses and Other Motivating Factors}\label{sec:motfac}

\citet[][hereafter K17]{kaib17} propose that the Fomalhaut triple star system system has been in a meta-stable bound state since its formation, devoid of catastrophic scattering events between Fomalhaut\,B and C such that we are not observing the system in a transient 
disruption state. The effect of the Galactic tide and passing field stars lead to a complex evolution of the eccentricity of Fomalhaut\,B's orbit around Fomalhaut\,A, such that periastron values low enough to excite the eccentricity of Fomalhaut\,A's belt may have been previously attained. K17 simulate the dynamics of the Fomalhaut system, starting with the stars at their present separations and with statistically generated orbital parameters. They evolve the system over 500 Myrs under the influence of the Galactic tide and passing field stars and classify a final state as a match to the real Fomalhaut system if the stellar separations are within 50\% of their current values. They find $\sim$7\% of their 2000 simulations end in a matching state, but that $\sim$51\% of systems passed through a matching state in the last 100 Myrs as systems oscillate between matching and unmatching. The systems that ended in a matching state are reintegrated from the initial conditions with an initially circular belt of 500 massless test particles between 127 and 143 au around A. They find 25\% of these systems end with an eccentricity between 0.04 and 1 for A's belt, due to close periastron passages of B. 
However, the standard deviations in longitude of pericentre and eccentricity of the test particles in these eccentric belts are significantly larger in the simulations than those derived by \citet{macgregor17} from ALMA observations of the belt. \citet{macgregor17} give their model particles a forced eccentricity and forced argument of periastron, as well as a proper eccentricity with a randomly distributed proper argument of periastron. These ranges of free eccentricities and periastron angles about the forced eccentricity result in a scatter of true eccentricities and pericenter angles for disc particles. The scatter in K17's model values for longitude of pericentre and eccentricity are both larger than in \citet{macgregor17}'s best fit model as well as being out of the range extrapolated from \citet{macgregor17}'s uncertainties.
Only 2 of the 135 simulated belts are matches to Fomalhaut's in all the above regards, namely median eccentricity, standard deviation in eccentricity and standard deviation in longitude of pericentre, and therefore apsidal alignment.
In all, this model is viable to explain the orbital configuration of the Fomalhaut system with $\sim$7\% of simulations resembling the current system after 500 Myrs; $\sim$25\% of these matching systems have close periastron passages of B that can excite the eccentricity of A's planetesimal belt, however only $\sim$1.5\% of the matching systems' belts (0.1\% of all simulations) are able to match A's in every regard.

\citet[][hereafter F17]{feng17} also modelled the Fomalhaut system under perturbations from the Galactic tide and stellar encounters. They initiate their models with the current relative stellar locations and integrate 500 Myrs backwards in time. C's orbit is classified as unstable if its orbital energy is larger than 0, i.e. it is unbound. They find that in all simulations C at least passes through an unbound state. In most models the separation between A and C only ever increases as the simulation progresses, but in a few percent of models C moves in and out of bound states and ends within 1 pc of A after the 500 Myrs. These are systems on meta-stable orbits, like those proposed by K17. These systems are termed `gravitational pairs` by F17 and likened to Cooper pairs in a superconductor, as the orbital binding energy of the system is comparable to the energy fluctuations from the Galactic tide and stellar encounters. As stable orbits are too rare and unstable orbits are too short lived, F17 conclude that A and C are likely one of these 'gravitational pairs'. They also find that in 20\% of models B comes within 400 au of A, thus likely being able to excite eccentricity in A's disc as shown by K17.

An alternative scenario is proposed by \citet[][hereafter S14]{shannon14}: A and C formed together as a binary from a single molecular cloud core which was then disrupted by the capture of B. To test this hypothesis S14 conduct N-body simulations with randomly sampled initial separations and eccentricities of the AC binary; B is generated at a random location within the Hill sphere of the system with a random velocity and eccentricity. The simulation is run for 500 Myr and stars are removed if they venture more than 2 pc from A. One thousand simulations were conducted and a match is defined by simultaneous separation of AB and AC within 0.5 -- 1.5 times their existing values. Over the 500 Myr run of the simulation 46\% have at least one period of matching; after 500 Myr 21\% of systems retain all three stars, of which 19\% were never matches. As these 19\% of systems remain to become unstable and may possibly match in the future, S14 estimate that in total 55\% to 60\% of systems will eventually pass through a Fomalhaut-like state. The matching state is temporary, on the scale of tens of Myrs, and often followed by an ejection, more often of C than B. To investigate the effects of such interactions on the discs around A and C, 50 further simulations were conducted with discs of 100 test particles placed around A and C randomly distributed within the then-known bounds of the two discs (127--143 au and 10--40 au, respectively). Of these, 38\% become a match over 500 Myrs. The discs are found to rise in eccentricity, with a high level of apsidal alignment, driven by secular interaction. Five of A's discs reach eccentricities of 0.02--0.5, reminiscent of the current Fomalhaut\,A system. 
Further close encounters can become disruptive and raise eccentricities to even higher values, but repeated close interactions are not guaranteed, allowing eccentricity to be preserved over the timescale of the matching state. 
For apsidal alignment to also be preserved, the timescale of differential procession would need to be longer than the timescale of the matching state.
The mean eccentricities of A and C's discs are correlated but show a strong scatter, Fomalhaut\,C disc eccentricities vary between $\sim$0.025 -- 0.75.

On the one hand, the hypothesis of S14 relies on a particular set of initial conditions, on the other, F17 and K17 make no statements on how the system would have formed. S14's models do have a $\sim$15\% success rate at describing both the current orbital configuration of the system and the morphology of A's eccentric belt. F17 find $\sim$1\% 
of their systems have a matching configuration and passages of B that could excite the eccentricity of A's belt and K17 find $\sim$2\% 
have a matching configuration and close passages of B; however K17 find these close passages only produce A-like disc morphologies in 6\% of cases. Na\"{\i}vely operating on these percentages alone it seems the S14 hypothesis is most likely, however the likelihood of the initial conditions arising in each of the three cases is not quantified. The S14 hypothesis may also require observing the system in a transient state just before a star is ejected, which is less likely than observing a system in a long-lived meta-stable state. 

These scenarios can be distinguished in several observational ways. An extremely precise measurement could be made of the individual stellar velocities 
to pin down the present orbital parameters, however, given the extremely large separations and the large timescale of the orbits and small orbital velocities involved this a very difficult task. Alternatively, K17, and by extension F17, predict A's belt to be significantly less apsidally aligned than S14 does, S14's apsidal prediction being more consistent with current observations. 
Another prediction of S14 is that the eccentricity of the belt of C should be correlated with that of A; if the eccentricity of C's belt were to be measured it could support or weaken S14's case. Such observations and measurements are presented and discussed in \S \ref{sec:ResAn} in this paper. The interaction proposed by S14 could also have driven planetary instabilities around C that later stir the disc or the collisional cascade directly, leading to its increased brightness and ease of detection, uncharacteristic of M-dwarf discs.

Disc detection rates are presently low around M-dwarfs: the Herschel DEBRIS survey detected just 2 debris discs from 94 M-dwarfs \citep[GJ\,581, Fomalhaut\,C;][respectively]{Lestrade12,Kennedy13} 
and a separate Herschel survey of M-dwarf planet hosts with greater sensitivity found 3 discs among 21 late-type stars \citep[18 M-dwarfs and 3 K-dwarfs;][]{Kennedy18a}. The key question remains whether true incidence rates for M-dwarf discs are similar to earlier type stars and it is the low luminosity of the host stars that limits their temperature and luminosity. \citet{Luppe20} find that this may be the case, thus requiring highly sensitive observations made at far-infrared/sub-millimetre wavelengths for detection. The alternative is that discs are indeed less common around these late type stars, perhaps due to effects that more significantly affect discs around low mass hosts such as stripping from stellar encounters \citep{Lestrade11} or photoevaporation of the primordial disc in cluster environments \citep{Adams04}. It is also possible that efficient planet formation around low mass stars could use up all the disc material, consistent with the increased planet occurrence rate measured for lower mass stars \citep[e.g.][]{Bonfils13,Dressing15,Mulders15}. 

Aside from increased stirring in S14's scenario, why else could Fomalhaut\,C have a detectable disc? Relative to a random selection of field M-dwarf ages, which span up to $\sim$10 Gyrs, Fomalhaut\,C is still young at 440 Myrs. Debris discs are typically found to be brightest when youngest, when their planetesimal belts have been depleted little by collisional evolution \citep{Decin03,Rieke05}, possibly explaining the presence of its bright disc. Fomalhaut\,C exists as one of the lowest mass nearby stars with a confirmed debris disc, and as one of just a handful of ALMA-detected M-dwarf debris discs, thus it will play an important role in our understanding of M-dwarf discs and the M-dwarf planet formation process.

\section{ALMA Observations}\label{sec:obsv}

We observed Fomalhaut\,C three times with ALMA in Band\,7 (0.87 mm, 345 GHz) from May 21st to June 6th 2018 under project 2017.1.00561.S. All observations used baselines ranging from 15 to 314 m and 48, 45 and 47 antennae respectively with an average precipitable water vapour of $\sim$0.7\,mm. The total on source observing duration was 102 minutes. J2148+0657 and J0006-0623 were used for pointing, bandpass and flux calibration. J2258-2758 was observed between individual target scans for time-varying atmospheric calibrations. Each pointing was updated for the proper motion of Fomalhaut\,C, however the proper motion over the two weeks ($\sim$ 0.017") between the first and last observation is negligible in comparison to the beam size ($\sim$ 1") and thus pointing differences are ignored when the observations are combined.

The spectral setup comprised four windows centered on 347.833, 335.791, 333.833 and 345.833 GHz with bandwidth 2\,GHz and 128 channels for all but the last, with width 1.875 GHz and 3840 channels of width 0.424\,km/s. The last window was used to search for CO gas via the J=3-2 emission line, which can be produced in planetesimal collisions and has been identified in the disc of Fomalhaut\,A \citep{matra17}. 

The raw data were calibrated with the provided ALMA pipeline script in CASA version 5.1.2-4 \citep{CASA}. To reduce the data volume the visibilities were averaged in 30s intervals and down to two channels per spectral window for the continuum imaging. All images were generated with the CLEAN algorithm in CASA.

\section{Results and Analysis}\label{sec:ResAn}
Given the relatively low signal to noise ratio (S/N) of the emission we carried out continuum imaging using natural weighting (equivalent to Briggs weighting with a robust parameter of 2) to preserve as much signal as possible and did not attempt self-calibration. We do not use a u-v taper as the disc is not well resolved radially, and one or more point sources within the primary beam begin to dominate the emission before disc structure is strengthened. This weighting gives a synthesised beam with a position angle (PA) of 83.16$^\circ$ and major and minor FWHM of 1.14" and 0.90" respectively, corresponding to 8.7 and 6.9\,au at a distance of 7.67\,pc. The standard deviation in the area around the disc is $\sigma = 17.5 \mu$Jy beam$^{-1}$ as identified by measurement from an annulus of sky exterior to the disc. This noise is for the most part uniform throughout the 4" radius centre of the image where the disc is detected, where the primary beam correction is $< 10\%$. 

\subsection{Initial continuum analysis}\label{sec:ICA}

\begin{figure}
        {\includegraphics[width=\columnwidth]
        {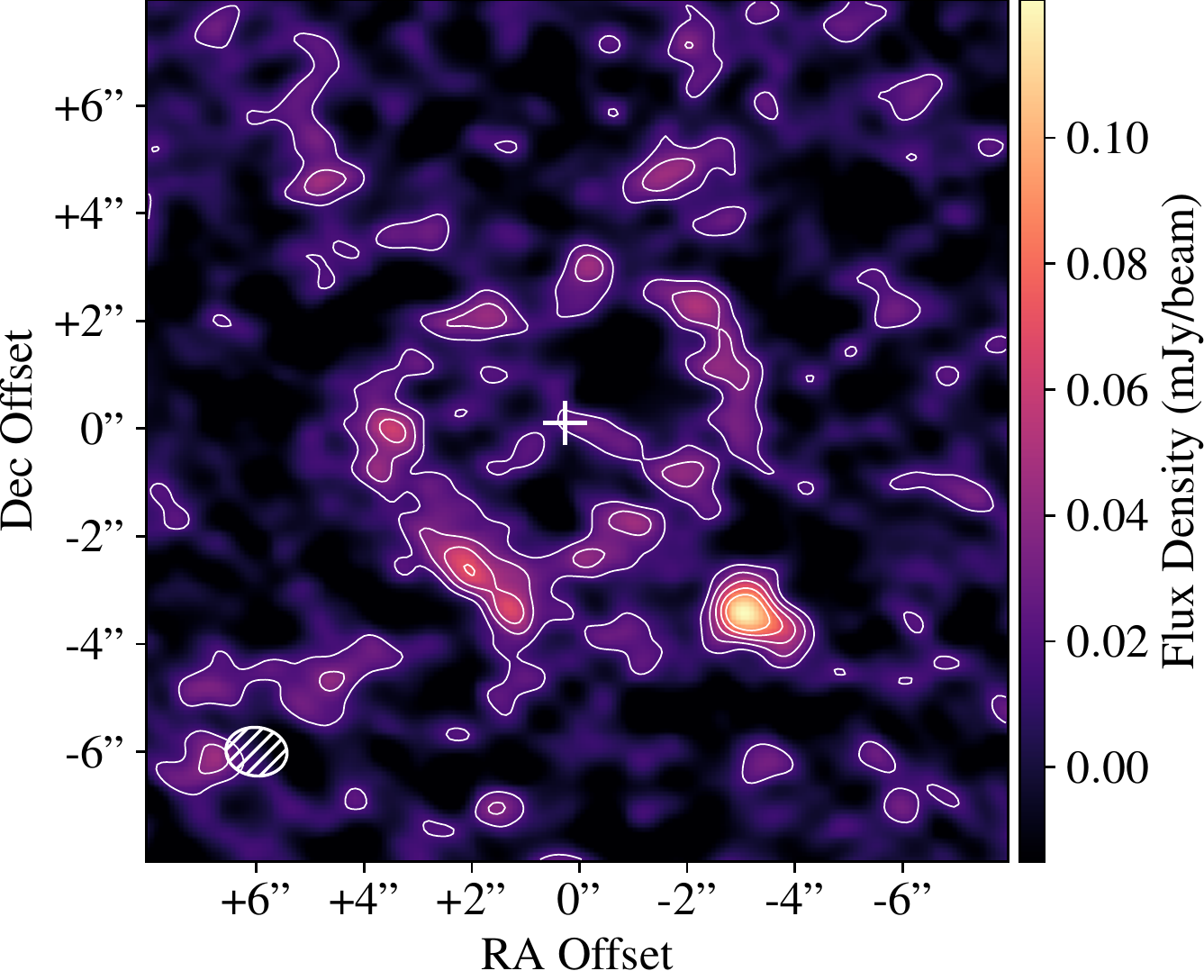}}
        \caption{\label{fig:FomCont}  Naturally-weighted clean image of the disc around Fomalhaut\,C. The ellipse in the lower left corner shows the beam size of 1.14$\times$0.90". The star is not detected. At a distance of 7.67\,pc the disc radius is 26.4\,au. Contours are drawn at $ 1, 2, 3, 4, 5  \sigma$ with $1\sigma = 17.5 \mu$Jy  beam$^{-1}$. The location of the star is marked with a $+$ at 342$^\circ$01'14.1" $-$24$^\circ$22'11.1". Zero offset is the ALMA image phase centre at 342$^\circ$01'13.8" $-$24$^\circ$22'11.2" (J2000).}
\end{figure}

We will present detailed modelling of the visibilities below, but we will discuss the CLEAN continuum image first for a qualitative introduction and outline.

Figure \ref{fig:FomCont} shows that Fomalhaut\,C's ring is not continuously detected at all azimuths, even to a $1\sigma$ level. Approximately half of the disc area is detected at a $2\sigma$ level with some peaks at 3 or $4\sigma$. Although the overall flux level is low, it is apparent that the flux constitutes an inclined ring, this is shown to be a consistent interpretation through the modelling. The disc width appears similar to the beam size, limiting the ring's radial and vertical extents to within $\sim$10\,au. 

\begin{figure}
        {\includegraphics[width=\columnwidth]
        {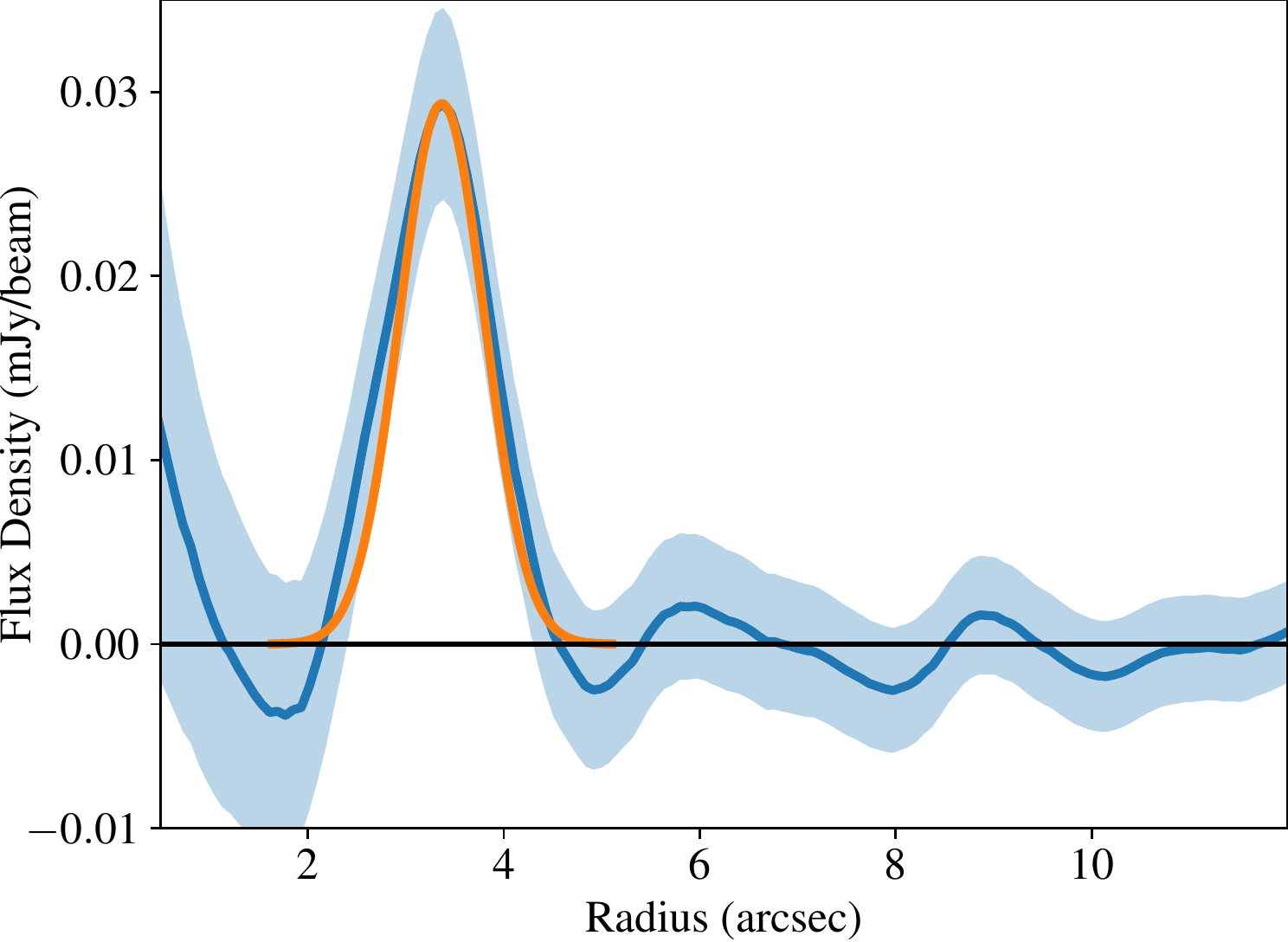}}
        \caption{\label{fig:RadProf} Deprojected radial profile of the disc around Fomalhaut\,C, computed from the CLEAN image within azimuthally averaged annuli. The shaded region denotes 1$\sigma$ uncertainties. A Gaussian with the same FWHM as the beam is plotted in orange at the peak radial flux.}
\end{figure}

Figure \ref{fig:RadProf} shows the disc's deprojected radial profile, assuming the disc PA and inclination found from the Gaussian Torus visibility modelling. Comparing the profile with a Gaussian with the same FWHM as the beam shows that the ring is strongly detected, but not radially resolved.

The disc appears brighter in the south-east quadrant with greater continuous >2$\sigma$ detection and larger amounts of >3$\sigma$ detection as well as the only $4\sigma$ peak within the disc at a PA of approximately 145$^{\circ}$ east from north. This variation is within expected noise fluctuations given a smooth disc but it is still investigated during modelling to explore whether or not this peak is a feature of the disc and whether or not its presence affects the fitted disc parameters. The peak S/N of the image is 7, located at a compact source in the south-west of the image 5" from the star. This external compact source we interpret as a background feature unassociated with the disc, but it is nevertheless included within the modelling.
The predicted stellar photospheric flux is 15\,\textmu Jy, consistent with there being no significant detection of the star and we find no evidence of stellar variability or flaring across the three observations. Preliminary fluxes can be taken from this image; $\sim0.7$\,mJy is measured from the image within an elliptical annulus containing the disc. A flux of $\sim0.2$ mJy is measured for the compact source in the south-west. Together these sum to $\sim0.9$\,mJy, which is consistent with the estimated disc flux of 1\,mJy initially extrapolated from Herschel measurements \citep[][\S\ref{sec:Herschel and Revised SED Model}]{Kennedy13}.

\subsection{Continuum Modelling} \label{sec:Continuum Modelling}

For a given set of parameters a rotation from sky coordinates to model coordinates (where the disc in is in the x-y plane) is calculated. A 3-dimensional model disc is then generated in the sky coordinates (RA, Dec, line of sight). Using the aforementioned rotation, the corresponding model coordinate is found for each pixel and the model is consulted to identify the model flux at each location. This disc model is then collapsed into a plane, creating a 2-dimensional image in the sky plane. This image then has any compact sources added as a symmetrical 2D Gaussian with a given centre, standard deviation and flux. The image is then Fourier transformed using the \textit{galario} package \citep{galario} and the u-v locations of the ALMA data are sampled to calculate the $\chi^2$ of the model given the data. We use the \textit{emcee} package \citep{emcee}, a Python implementation of the Markov Chain Monte Carlo method, to explore the posterior probability distributions of our model parameters in order to derive the best-fitting model. The models are initiated near the optimal solutions indicated by previous test model iterations. We use 5000 steps, with the first 3500 being discarded based on the estimated auto-correlation lengths. For the runs we use 200 walkers and we verify upon completion that all chains have converged.

Three distinct models were implemented in order to investigate the nature of the over-brightness in the south-east of the disc and its effect on model fitting. The external south-west compact source is marginally resolved and so treated as an azimuthally symmetric two dimensional Gaussian source and is included in all models.

The \textbf{Torus} model serves to model the disc alone as a comparison for the later models, here the south-east over-brightness constitutes simply a noise peak. The \textbf{Torus + Asymmetry} model treats the over-brightness as a feature of the disc which is thus contained within it, representing a local over-density of dust within the disc: perhaps a dust trap, pressure maximum, or recent collisional event. The \textbf{Torus + Point Source} model treats the over-brightness as unrelated to the disc but as a real feature of the image, possibly representing a background galaxy, to be accounted for so as to not affect the parameters of the disc when fitting.

Thus all disc models share these common parameters: the disc's total flux $F$, the disc average radius $r_0$, the disc's Gaussian scale height $\sigma_h$ (defined by angular elevation from the disc midplane), and scale width $\sigma_r$ (defined by radial distance from centre of disc), the disc position angle $PA$ (defined as east from north), the disc inclination $I$, the sky offset of the disc centre from the phase centre $x_0, y_0$; and the radial distance from the centre of the image of the external compact source in the south-west, the compact source's azimuthal angle in the image (measured east from north), the compact source's Gaussian scale width and the compact source's brightness. We find the phase-centre of the ALMA observations, and thus the image centre, to be slightly offset from the Gaia DR2 \citep{Gaia1,Gaia2} location of the star at the time of observation and have corrected all further mention of disc offsets for this such that disc offset is always measured from the stellar location.

These Gaia corrected $x_0, y_0$ offsets are on the plane of the sky, but any physical offset will also have some extent into (or out of) the plane of the sky. By assuming the offset is in the plane of the disc, this can be calculated using the sky offsets, position angle and inclination of the disc. 

This offset is calculated for every walker at every step to also produce a $z$ offset that is combined with the sky $x$ (RA) and $y$ (Dec) offsets to derive the total offset of a given model. This total offset is then divided by that individual model's disc radius to calculate an eccentricity. The eccentricity upper limit presented in Table\,\ref{tab:modelresults} is derived from the one sided 3$\sigma$ value of the final distribution of model eccentricities. This eccentricity upper limit also factors in the ALMA pointing uncertainty. The level of eccentricity derived in our models is small enough to still be well approximated by an offset circular disc, so a physically eccentric disc model is never explicitly used or needed to fit the disc.

\begin{table*}
\centering
\renewcommand{\arraystretch}{1.5}
\caption{Median disc parameters, $\Delta \chi ^2$ and $\Delta$BIC values for Torus, Torus with Asymmetry and Torus with Point Source models. Uncertainties are the 16th and 84th percentiles. We find no significant degeneracies between model parameters. Offsets are measured from disc model centre to Gaia DR2 location of the star. Upper limits are one sided at 3$\sigma$, i.e. the 0.996 quantile. The eccentricity upper limit includes the ALMA pointing uncertainty and the $z$ offset. $\Delta \chi ^2$ and $\Delta$BIC values relative to Gaussian Torus model with values 3278000.7 and 3278180.7 respectively, calculated from a model produced using the median parameters. }
\label{tab:modelresults}
\begin{tabular}{lcccc}
\hline
Parameter & Torus & Torus + Asymmetry & Torus + Point Source \\
\hline                                     
RA Offset (")                 & $0.04^{+0.08}_{-0.08} $   & $0.05^{+0.08}_{-0.09}$       & $0.07^{+0.08}_{-0.08}$  \\
Dec Offset (")                & $-0.07^{+0.08}_{-0.09} $   & $-0.07^{+0.09}_{-0.09}$       & $-0.02^{+0.07}_{-0.07}$ \\
Eccentricity                  & $0.04^{+0.03}_{-0.02}$  &   $0.04^{+0.03}_{-0.02}$      & $0.04^{+0.02}_{-0.02}$\\
Eccentricity $3\sigma$ Upper Limit& $0.14$  &   $0.14$      & $0.12$\\
Inclination ($^{\circ}$)      & $43^{+3}_{-4}$      & $42^{+4}_{-4}$            & $44^{+3}_{-3}$          \\
PA ($^{\circ}$)               & $-59^{+7}_{-6}$   & $-58^{+7}_{-6}$          & $-63^{+6}_{-5}$         \\
Disc Flux (mJy)               & $0.9^{+0.1}_{-0.1}$    & $0.9^{+0.1}_{-0.1}$    & $0.8^{+0.1}_{-0.1}$          \\
Radius (au)                    & $26.5^{+0.5}_{-0.5}$    & $26.4^{+0.6}_{-0.7}$            & $26.4^{+0.6}_{-0.6}$            \\
Scale Width (")               & $0.11^{+0.15}_{-0.08}$  & $0.14^{+0.16}_{-0.09}$          & $0.11^{+0.13}_{-0.07}$      \\
Scale Width $3\sigma$ Upper Limit (")& $0.6$  & $0.6$          & $0.6$      \\
Scale Height (Rad)            & $0.20^{+0.08}_{-0.08}$  & $0.22^{+0.10}_{-0.09}$       & $0.15^{+0.08}_{-0.07}$             \\ 
Scale Height $3\sigma$ Upper Limit (Rad)& $0.5$  & $0.7$     & $0.4$             \\ 
$N_{\text{Parameters}}$       & $12$                & $15$        &     $16$   \\  
$\Delta\chi^2$                & $0$         &  $-2$  &    $-20$    \\ 
$\Delta$BIC                   & $0$         & $+42$    &  $+40$         \\
   
\hline
\end{tabular}

\end{table*}

\subsection{Gaussian Torus} \label{sec:Gaussian Torus}

\begin{figure*}
        \centering
        \includegraphics[width=0.336\linewidth,height=0.325\linewidth]{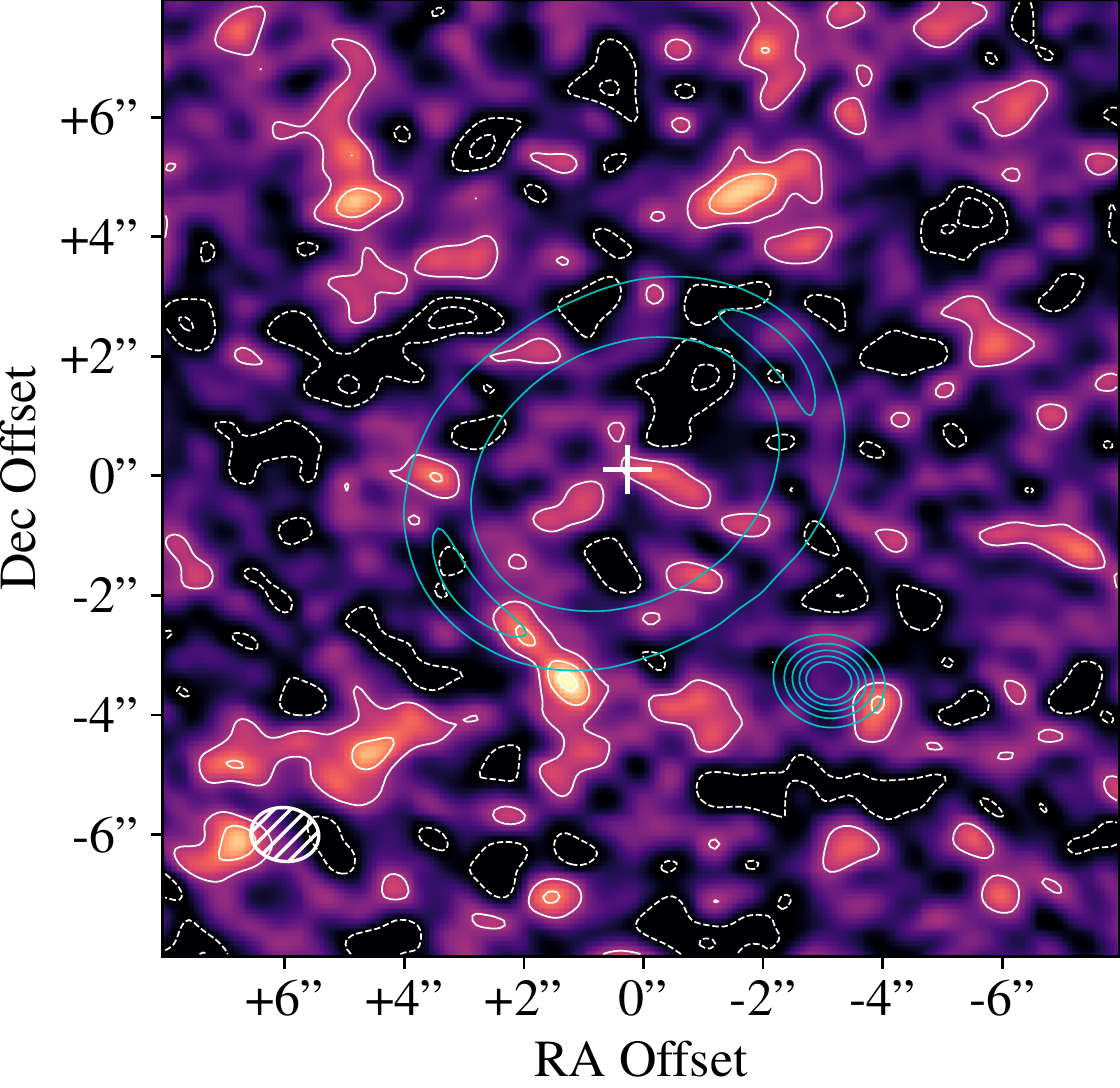}
        \hfill
        \includegraphics[width=0.287\linewidth,height=0.325\linewidth]{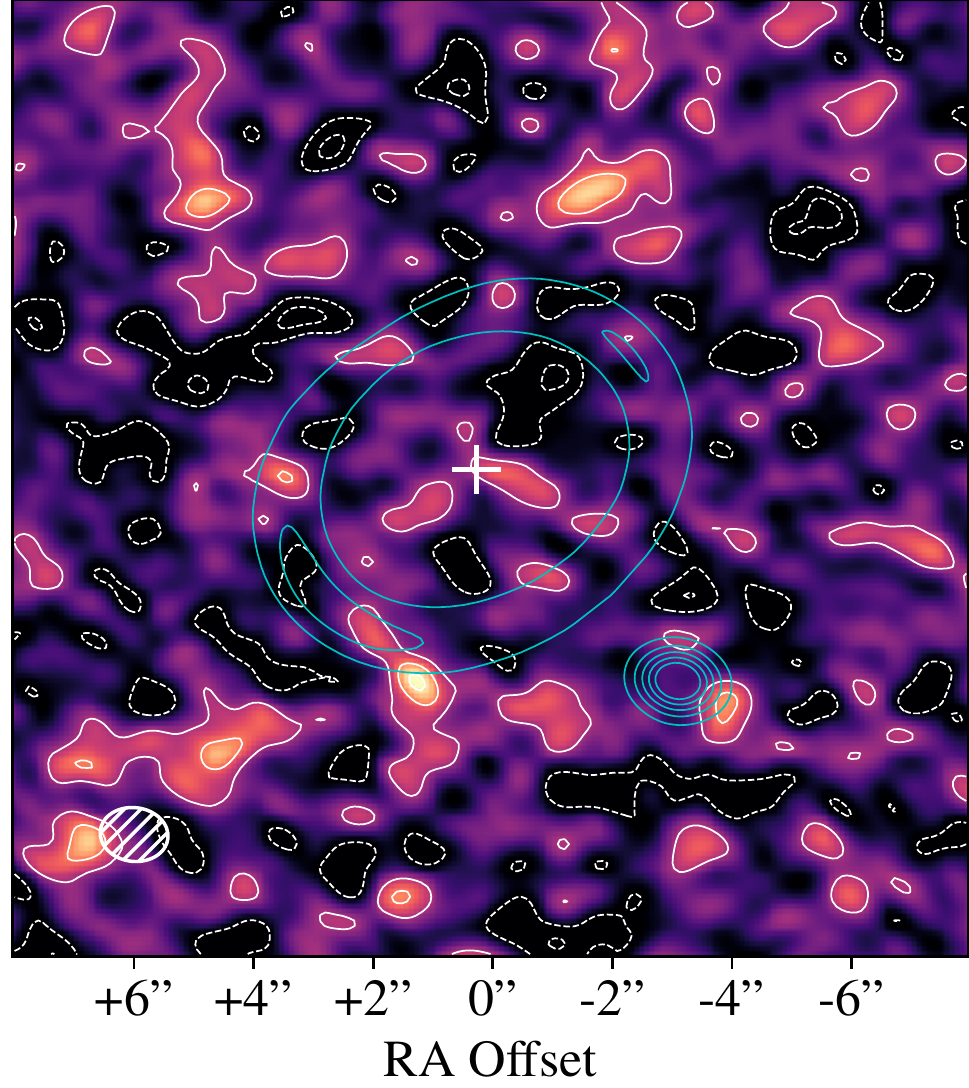}
        \hfill
        \includegraphics[width=0.353\linewidth,height=0.325\linewidth]{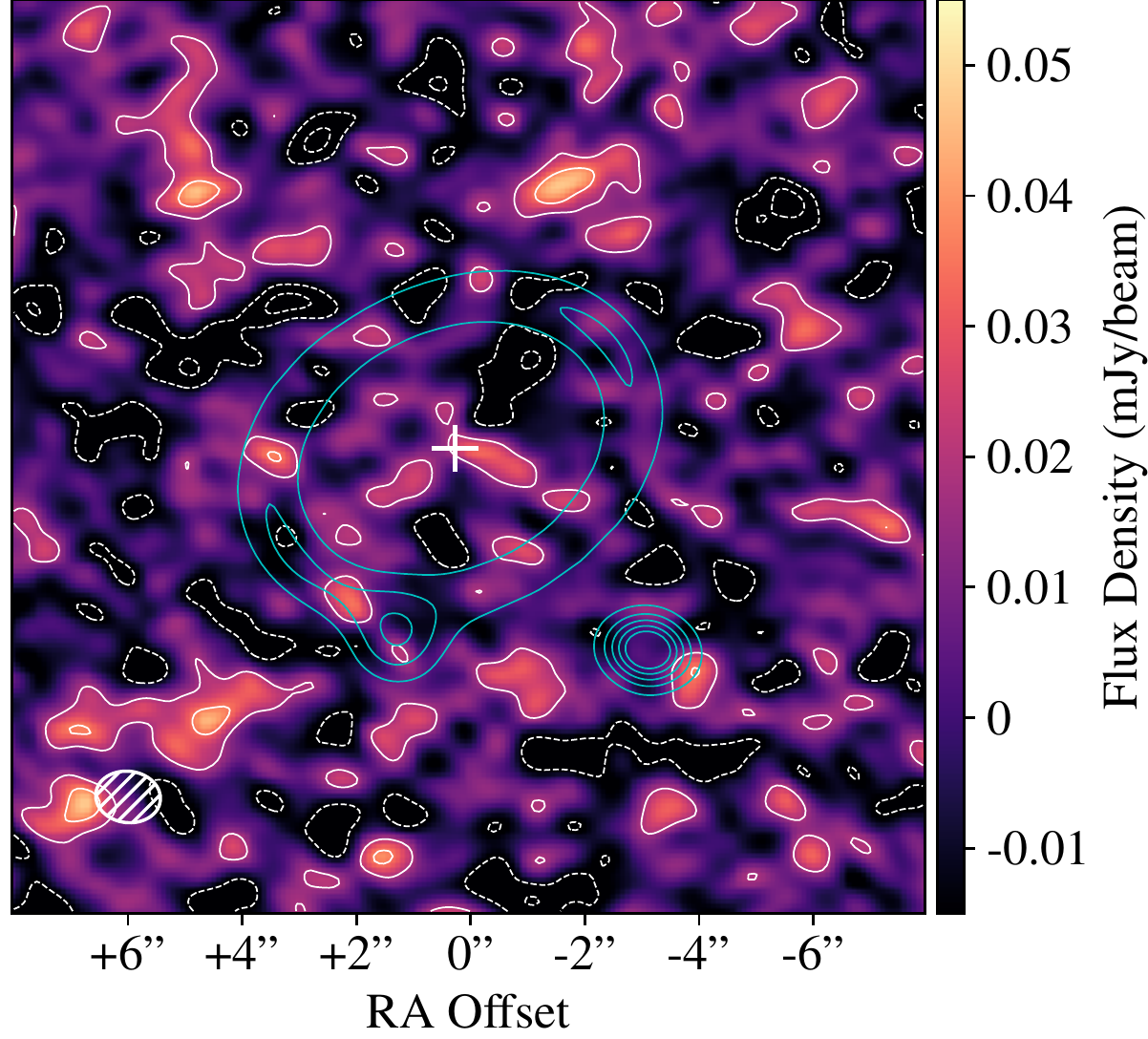}
    \caption{\label{fig:allres}Naturally-weighted dirty images of the residuals after subtracting the individual models. Left: Gaussian torus model; Middle: asymmetric torus model; Right: torus with point source model. Cyan contours show the models and white contours show the residuals at $-1, -2, 1, 2, 3, 4, 5  \sigma$. The location of the star is marked with a $+$. Zero offset is the ALMA image phase centre at 342$^\circ$01'04.9" $-$24$^\circ$22'11.2" (J2000).}
\end{figure*}

This model is the simplest and serves as our reference point. The best-fitting parameters are shown together with the other models in Table\,\ref{tab:modelresults} and a dirty image of the residuals after subtracting the visibilities of the best-fitting model is shown in Figure \ref{fig:allres} left. No discernible structure remains in the image showing that a azimuthally symmetric ring is a good representation of the data. The compact source in the south-west, outside of the disc, is also very well accounted for by the model. Using the medians of the posterior parameter distributions we calculate a $\chi^2$ value of 3278000.7 for this model; this can be compared to the values of the other models to quantify their relative goodness of fit. We also include the relative Bayesian Information Criterion \citep[BIC;][]{Schwarz78}, which tests whether the difference in $\chi^2$ values between models is significant by penalising models with extra fitted parameters, as can be seen in its definition: $BIC = \chi^2 + N_{\rm Parameters}\times\ln{N_{\rm dof}}$. As the number of visibilities ($N_{\rm dof} = 2\times N_{\rm vis} =2 \times 1639088$) being fitted is very large, there is a large penalty on the less simple models. A difference in BIC greater than six is considered 'strong' evidence that the lower valued model is preferred and a difference greater than ten is considered 'decisive' \citep{Kass95}.

In Figure \ref{fig:allres} left we can see that after the subtraction of the disc model the peak in the south-east remains at a significance of 3$\sigma$ with a larger 2$\sigma$ extent and a total flux of about 60\,$\mu$Jy. The 3$\sigma$ peak is located just outside of the disc's main emitting region, and the 1$\sigma$ extent reaches significantly into the disc, culminating in a 2$\sigma$ peak. There do exist multiple other 2$\sigma$ peaks within the image, but only one other 3$\sigma$ peak within the FWHM of the primary beam. This region is also the only one co-located with the known disc emission and could thus affect the fitting or be physically associated. Residuals do remain in roughly this location across the 3 separate nights of observations, however given the even lower signal to noise of the individual nights they do not offer much information when not combined.

\begin{figure}
        {\includegraphics[width=\columnwidth]
        {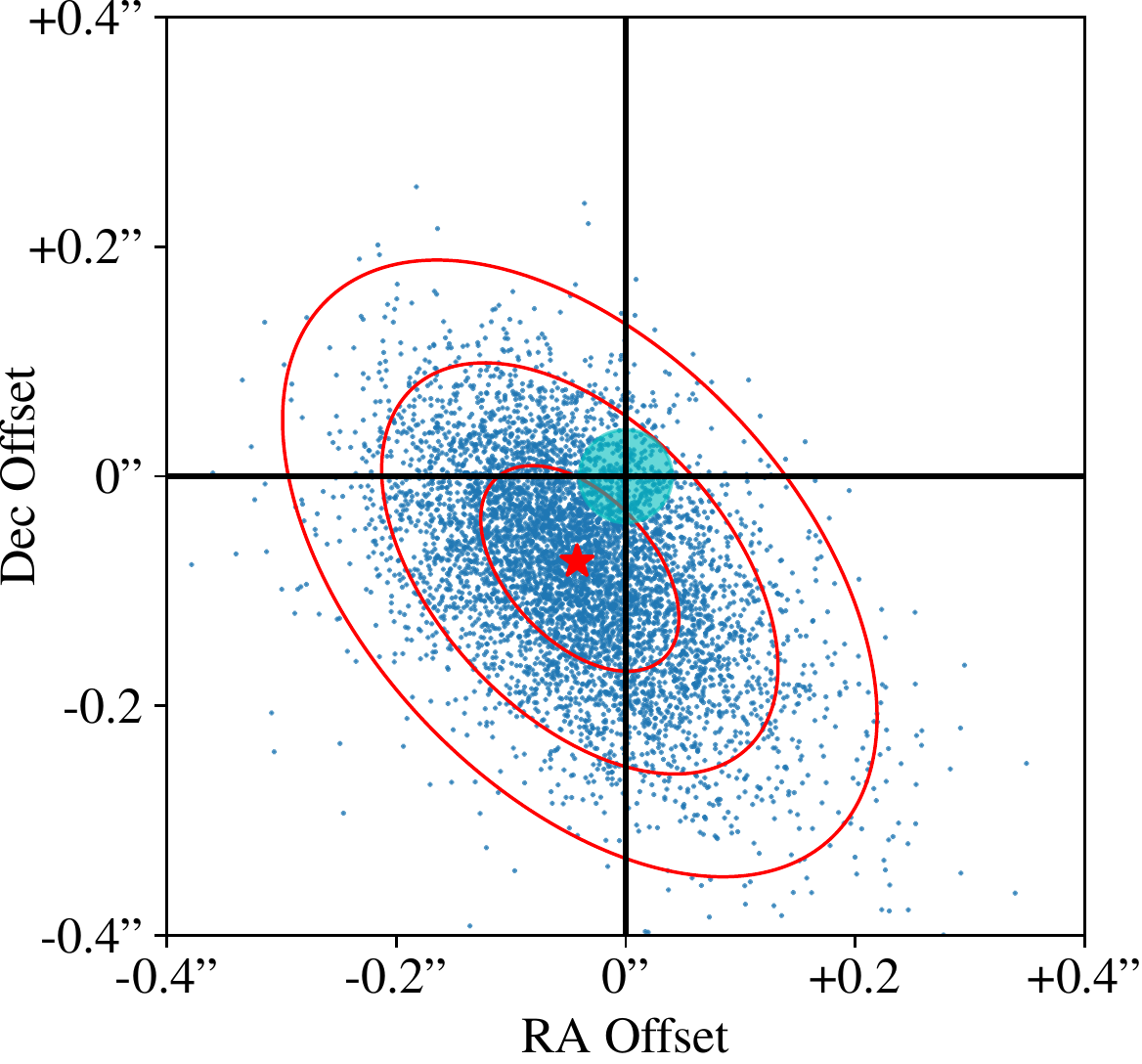}}
        \caption{\label{fig:Offsets} Distribution of offsets from \textit{emcee} for the torus model. The stellar location is at the origin. Blue points are individual model disc centre offsets from each walker at every 50th step after initial discarding. The red star denotes the median offset and the successive red ellipses contain 68, 95 and 99.7 \% of the offsets respectively. The blue shaded region is the 1$\sigma$ ALMA absolute pointing precision. The Gaia stellar location uncertainty would be too small to be seen.}
\end{figure}

We can draw some preliminary conclusions from this basic model. A disc does fit the data well, with a moderate inclination of about $40^{\circ}$ and a PA of about $-60^{\circ}$. The radius is well defined at 26.5 au and the disc flux is around $0.9$\,mJy. The south-west external compact source has a flux of around $0.2$\,mJy. As their posterior distributions are consistent with zero, we take the scale width and height to be unresolved and conclude only upper limits are obtainable. The model does find that the disc centre is offset from the stellar location, by $0.15\pm 0.09"$. 
This value is the median of the total three dimensional offset distribution found by the modelling, and so is not equal to a quadrature combined two dimensional sky offset calculated from the median RA and Dec offsets presented in Table\,\ref{tab:modelresults}.
The uncertainty in this offset value is large, much larger than uncertainty of the Gaia DR2 location (0.00034") and the pointing uncertainty of ALMA (0.0405", here taken as 5\% of the beam FWHM) as shown in Figure \ref{fig:Offsets}. Figure\,\ref{fig:Offsets} shows the two dimensional distribution of offsets as well as the ALMA pointing uncertainty. The offset distribution overlaps with zero between 1 and 2$\sigma$, and overlaps with the ALMA pointing offset uncertainty at 1$\sigma$. The elliptical distribution of the offsets is to be expected, as there is less spread along the major axis of the disc where the S/N is highest, allowing for more precise fitting. 

The median offset is $1.2\pm 0.7$\,au, corresponding to an eccentricity of $0.04 \pm 0.02$. Due to the large uncertainty we do not take this result to be significant evidence of an offset, and we instead place a 3$\sigma$ upper limit on the eccentricity of 0.14. Our constraint on the eccentricity is limited by both the low S/N of the data and the ALMA pointing precision at 1" spatial resolution.

With a model flux value for the point source in the south-west we can make an estimate for the likelihood of such a background galaxy being present within the image. We will compare with the 1.2 mm galaxy number counts of \citet{Aravena16} by converting our ALMA Band\,7 870 \textmu m flux to a 1.2 mm equivalent. \citet{Aravena16} provide a conversion for flux $S$ between different bands: $S_{1.2mm}=0.4S_{870\mu m}$. Our 870 \textmu m flux of 0.2 mJy is then equivalent to a 1.2 mm flux of 0.08 mJy. From Table\,2 of \citet{Aravena16} we estimate that there are 23,700 sources per square degree with a flux greater than 0.077 mJy. The probability of finding at least one galaxy brighter than this within a central 8 arcsecond radius, approximately the area of interest around the disk as shown in Figure\,\ref{fig:FomCont}, is then around 30 per cent. And so, the simplest explanation is that the south-west point source is  background galaxy.

\subsection{Gaussian Torus with Asymmetry}

This model treats the south-east over-brightness as a component of the disc in the form of a 3-dimensional Gaussian blob embedded within it, representing a substructure. This adds three extra parameters to the model: the blob's azimuth, the blob's azimuthal extent, and the blob's brightness. The minimum azimuthal extent is taken as the beam size (i.e. 17$^{\circ}$), and the maximum azimuthal extent is taken as a quarter of the disc (i.e. 90$^{\circ}$). The blob is centred within the disc and shares the disc's width.
The disc flux value is the sum of the flux from the main disc and the blob contained within. 

From examination of the residuals for this model, shown in Figure\,\ref{fig:allres} middle, we can see that the in-disc residual in the south-east is reduced in size compared to Figure \ref{fig:allres} left and no longer reaches $2\sigma$. Compared to the torus model, we can also see that the north-west ansa has decreased in flux, showing that the disc's general flux has decreased, with the asymmetry taking up the extra flux needed at the south-east ansa. The asymmetry itself makes up $10^{+16}_{-7}$ per cent of the flux of the disc, for a disc flux of 0.9 mJy this corresponds to $0.09^{+0.14}_{-0.06}$ mJy. The distribution is consistent with zero flux showing that the asymmetry is not required to replicate the data. The asymmetry is centred $3\pm 25^{\circ}$ counter-clockwise from the south-east ansa of the disc and has a Gaussian scale azimuth of $34\pm 16^{\circ}$. 

This fit slightly decreases the $\chi^2$ value, but has a large increase in BIC. This increase in the BIC shows that the model does not justify the inclusion of extra parameters, consistent with the flux of the blob being consistent with zero. Most of the disc parameters remain very similar to the torus model with the only notable change in parameter value being a slight increase in the $3\sigma$ upper limit in scale height.

\subsection{Gaussian Torus with Point Source}

This model treats the south-east over-brightness as a background compact source, similar to the external compact source in the south-west. It also adds four extra parameters to the basic model, the radial distance from the centre of the image of the point source, its azimuthal angle in the image (measured east from north), its Gaussian scale width and its brightness. The fitting is also restricted such that the point source can only reside within the vicinity of the disc in the south-east quadrant.

Upon inspecting the residuals for this model in Figure\,\ref{fig:allres} right it can be seen that not even a $2\sigma$ contour remains in the south-east region of the disc. The $\chi^2$ of this fit is also significantly less than the other two models, being 20 less than the basic Gaussian torus model showing that it fits the data best. However, the flux of the compact source is consistent with zero and the BIC is still significantly larger than for the basic torus model, meaning that the inclusion of extra parameters is not justified by the decrease in $\chi^2$. Whether or not the over-brightness truly is the result of a background source is less important; what this model allows us to consider is how the disc is fit without its influence. In this model the over-brightness point source accounts for $0.1$\,mJy of flux, and the rest of the disc possesses just $0.8$\,mJy. 
The disc flux is consistent with the flux of the previous models but these values show that the model disc fluxes could be inflated if the south-east point source is real and not associated with the disc.
Again, while within uncertainty, the PA of the disc has relaxed to $63^\circ\pm 6$ as opposed to the previous two models' $58^\circ\pm 7$. This is not a significant effect but may be a sign that the fitting was attempting to align the south-eastern ansa of the disc with the over-brightness to account for it. 
The reduction in Dec offset and slight increase in RA offset could also be attributed to a similar effect. With the addition of the point source, the disc has shifted to the north-east, moving the ansa away from the over-brightness. But, although the direction of the offset has shifted, the magnitude has not been significantly reduced as can be seen from the derived eccentricity and eccentricity upper limit.
We also see a return to a more moderate scale width than the previous model and a slight reduction in scale height. The flux density distribution of the two compact sources is unknown, but their contributions at shorter wavelengths could contaminate the disc flux from the Herschel data, this possibility is explored in \S\ref{sec:Herschel and Revised SED Model}.

\subsection{Continuum Modelling summary}
\label{sec:ModSum}
In summary, the debris disc around Fomalhaut\,C is detected and resolved with ALMA, and the radius and orientation are well constrained. A Gaussian torus represents the dust ring well, but the radial and vertical scale heights are unresolved, with only upper limits available. Disc parameters are consistent with each other between the different models, but only the torus with point source model does not leave $2\sigma$ residuals in or near the south-east sector of the disc. The basic torus model has the lowest BIC and thus is the preferred model.
As the basic torus model's residuals leave the over-brightness mostly outside of the disc's bound and as the asymmetric disc model failed to find a significantly better fit, it can be concluded that the south-east over-brightness is not likely associated with the system. 
That the torus with point source model did not find a significantly better fit than the other models implies that the over-brightness is most likely just a noise peak. It could be a background object, but our BIC values show that there is not enough significance to conclude such. The similarity across modelling results finds that this feature does not significantly affect the fitted disc parameters. If real, observations at a later epoch would be able to confirm the nature of the point source if it does not share the proper motion of the star. 

A small offset of the disc centre from the star is consistent across all models, but is not significant. In all our models, the distribution of offsets retrieved from \textit{emcee} overlaps with zero between 1 and 2$\sigma$, and overlaps with the ALMA pointing offset uncertainty.

\subsection{CO Non-Detection}\label{sec:CO}

The spectral setup of the ALMA observations was designed to allow a search for CO gas produced in collisions of planetesimals that are rich in volatiles via the J=3-2 emission line. After subtracting the continuum emission, visual inspection revealed no clear signal in both the dirty cube and a moment-0 map produced by summing pixels across the channels in the velocity range where gas is expected. 

\begin{figure}
        {\includegraphics[width=\columnwidth]
        {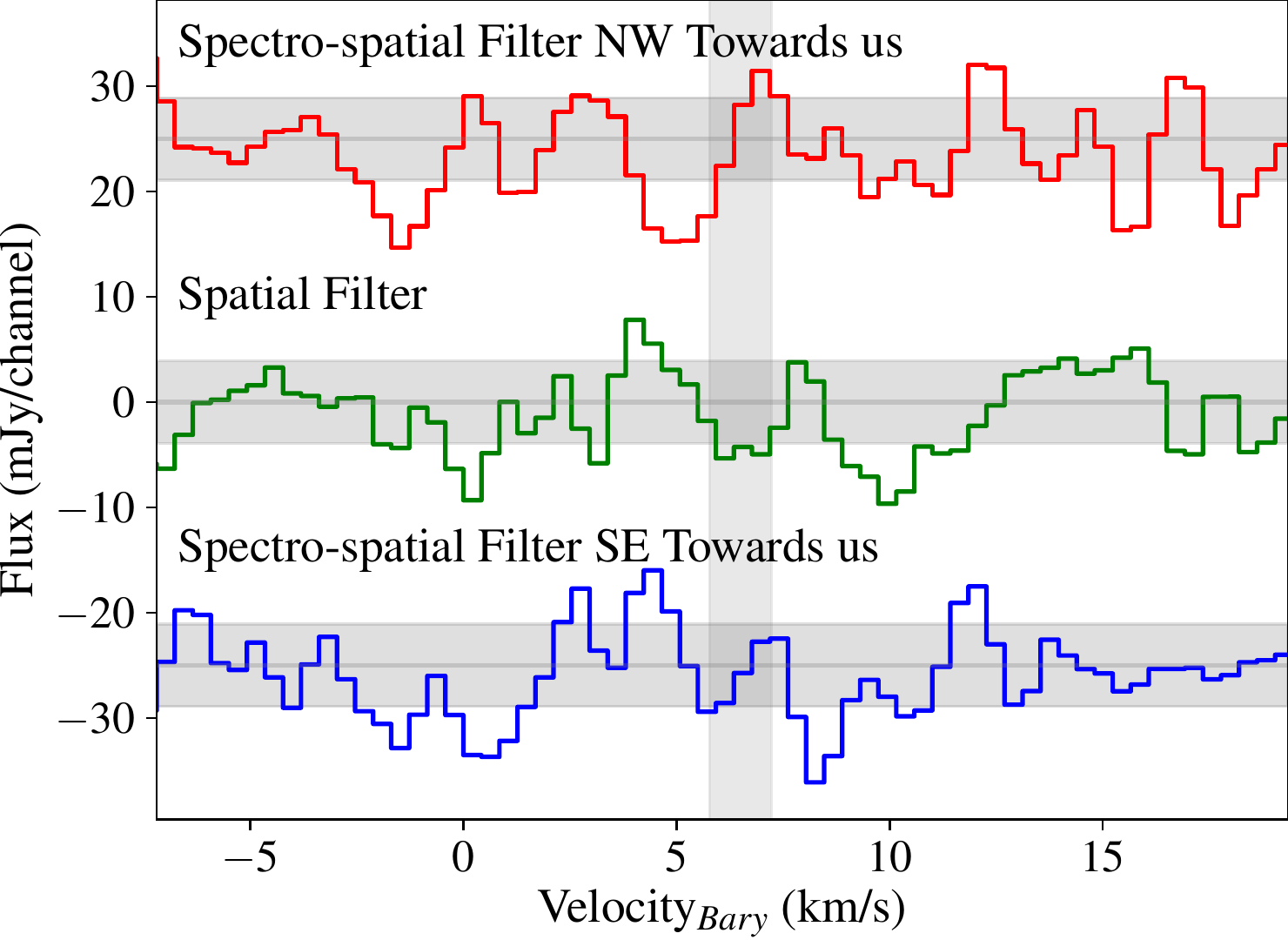}}
        \caption{\label{fig:CO}  CO J=3-2 spectra for the debris disc around Fomalhaut\,C using spatial and spectro-spatial filtering techniques. The centre spectrum (green) is filtered only by the bounds of the disc as per our Gaussian Torus model. The top and bottom spectra have had disc pixels shifted within the data cube by their expected Keplerian velocities, and are vertically displaced for graphical clarity. As there are two possible rotations of the disc with either ansa rotating towards us there are two possible shifts. There is no significant signal in any spectrum. Horizontal shaded regions denote the 1$\sigma$ uncertainty of the spectrum taken over a larger range of velocities after subtraction of a second order polynomial background. The vertical shaded region denotes the expected centre of the signal at the 6.5$\pm 0.5$\,km\,s$^{-1}$ stellar radial velocity.}
\end{figure}

To enhance the signal of potential CO in the system we also employed the spectro-spatial filtering approach as described in \citet{Matra15} under the assumption that any CO present would be co-located with the dust. In this method pixels are spectrally shifted within the data cube to account for the expected radial velocities from the Keplerian motion within the disc, here we assume a stellar mass for Fomalhaut\,C of 0.18\,M$_\odot$ \citep{Pecaut13}. We use the torus model from \S \ref{sec:Gaussian Torus} as a spatial filter, masking all pixels that are not co-located with model continuum emission that reaches at least 10\% of the peak model flux. Figure \ref{fig:CO} shows the corresponding spectra for the spatial filter alone and the spectro-spatial filter assuming either the north-west or the south-east ansa is rotating towards us. No signal is discernible in any of the produced spectra and so we calculate a 3$\sigma$ detection limit. With the application of the spectro-spatial filtering and with channel widths of 0.424\,km/s we calculate a 3$\sigma$ upper limit on the CO flux of 16 mJy\,km\,s$^{-1}$. This limit accounts for both the 10\% flux calibration uncertainty from ALMA and the correlation of adjacent channels.

A direct flux comparison of the CO non-detection for Fomalhaut\,C to the CO detection for Fomalhaut\,A is not straightforward, as the latter observations were not of the CO J=3-2 transmission line, but of the CO J=2-1 transmission line with ALMA Band 6 \citep{matra17}, and the CO excitation is uncertain. We might compare to the initial ALMA Band\,7 observation of Fomalhaut\,A \citep{Matra15} in which a flux limit of 160 mJy\,km\,s$^{-1}$ was calculated, however this was an ALMA Cycle 0 observation and the continuum sensitivity was also 3.5 times lower than in our observation. An appropriate example against which to compare Fomalhaut\,C is provided through the M-dwarf TWA\,7 \citep{Matra19} for which CO J=3-2 was detected with a integrated flux of 91\,$\pm$\,20\,mJy\,km\,s$^{-1}$ at a distance of 34 pc. 

To set an approximate constraint on the CO+CO$_2$ mass fraction of the planetesimals, we make a simple comparison of the collisional mass loss rate and flux limit with those of TWA\,7. Following the prescription set out in the appendix of \citet{matra17}, we compute the mass loss rate for Fomalhaut\,C's smallest grains of $\dot M_{D_{\rm min}}$. The minimum grain size, $D_{\rm min}$, is an unknown here in the regime of stellar wind dominated grain removal and we do not have enough short wavelength data to retrieve an estimate from the flux density distribution as in \citet{Matra19}. The minimum grain size for TWA\,7 was found to be 0.1 \textmu m and using this number as a fiducial value for Fomalhaut\,C we get $ \dot M_{D_{\rm min}} = 6\times 10^{-5} $ $ M_\oplus$\,Myr$^{-1}$. For comparison the value is $3 \times 10^{-3} M_\oplus$\,Myr$^{-1}$ for TWA\,7. The CO+CO$_2$ mass (which photodissociates in time $t_{\rm phd}$) is estimated as
\begin{equation}
M_{\rm CO+CO_2} = t_{\rm phd} \frac{f_{CO+CO_2}}{1-f_{CO+CO_2}} \dot M_{D_{\rm min}} \, ,
\end{equation}
where $f_{CO+CO_2}$ is the fraction of planetesimal mass in CO and CO$_2$ ice \citep{matra17}. Thus, if we assume the same CO excitation and lifetime for Fomalhaut\,C as for TWA\,7, and that the observed CO flux is proportional to the CO mass, then it is only the difference in mass loss rates and planetesimal CO+CO$_2$ fraction that changes the observed CO flux. TWA\,7 has a 50$\times$ higher mass loss rate and CO flux 5$\times$ higher than our upper limit, but Fomalhaut\,C is 4.4$\times$ closer, so with Fomalhaut\,C we could have detected CO at half of TWA\,7's observed level for the same CO+CO$_2$ fraction. Thus, $f_{CO+CO_2}$ is constrained to be at least $\sim$2$\times$ lower than for TWA\,7. The ice fraction for the planetesimals of TWA\,7 was found to be $\geq70\%$, so for our assumptions the non-detection does not appear to be particularly constraining compared to the $\lesssim$10\% fractions observed in the Solar system \citep{leroy15}. We can now also compare our ice fraction constraint with Fomalhaut\,A's ice fraction of 4.6\%--76\% \citep{matra17} to find that the two are consistent.

\subsection{Herschel/PACS Modelling and Revised SED Model} \label{sec:Herschel and Revised SED Model}
\begin{figure}
        {\includegraphics[width=\columnwidth]
        {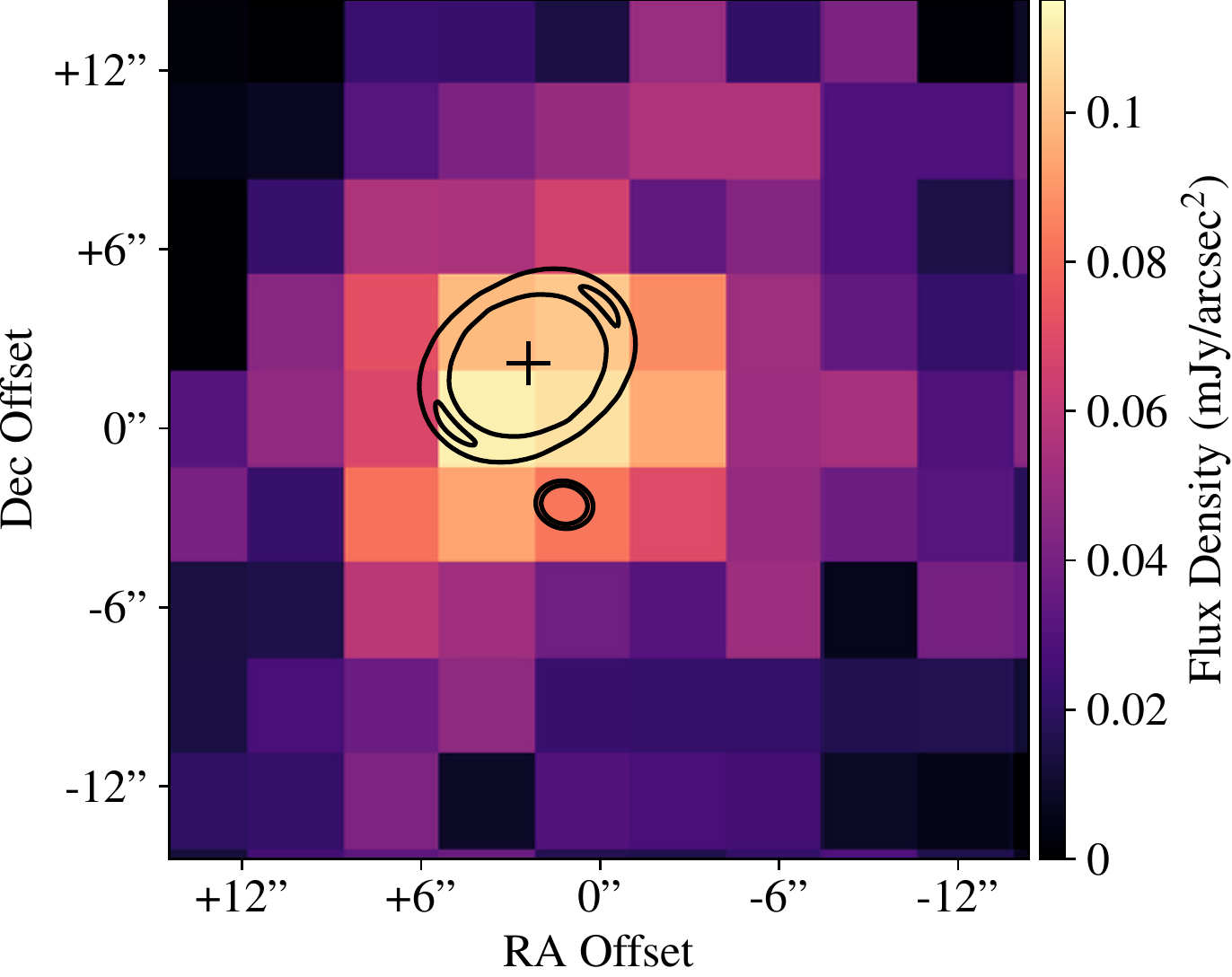}}
        \caption{\label{fig:Herschel} Herschel 160 \textmu m PACS detection of the Fomalhaut\,C disc. In black are contours of $1, 2,\sigma$ from the ALMA best fit torus model, assuming zero proper motion of the external compact source appropriate for a background galaxy. The Gaia DR2 location of the star is marked with a $+$. Zero offset is at 342$^\circ$01'09.6" $-$24$^\circ$22'12.1" (J2000). Our best fit pointing correction of +0.54" RA, +0.33" Dec has been applied to the Herschel image.}
\end{figure}

With the additional knowledge of the Fomalhaut\,C disc's geometry, and of the presence of a nearby background compact source, it is worth re-analysing the Herschel data \citep{Kennedy13} to see if new information can be gleaned, or to see if the background source partially contaminated the original detection. Total contamination of the original detection is highly improbable as the chance of detecting a debris disc around a randomly chosen M\,star is already very low. We use the level 2.5 data product 160 \textmu m PACS image from November 2011 (Observation IDs 1342231937, 1342231938; Observing Day 906) as our data for model comparison. For reference, we show our best fit ALMA torus model in contours over the Herschel detection in Figure \ref{fig:Herschel}. 

A similar approach was taken to modelling the Herschel data as was taken for the ALMA data as described in \S \ref{sec:Continuum Modelling}. A model is generated using the median disc parameters of the Gaussian torus model from \S\ref{sec:Gaussian Torus}. The disc model's stellar location is centred on the Gaia DR2 location of Fomalhaut\,C at the time of the Herschel observation, however the compact source is assumed to be in the background and thus its position is not corrected for proper motion between the dates of the Herschel and ALMA observations. The entire model is then allowed to be offset from the centre to account for the imprecise Herschel pointing, with a Gaussian prior set on the offset using the absolute pointing performance of 1.12" at 1$\sigma$ provided by ESA within the observing date range of the observation\footnote{\url{https://www.cosmos.esa.int/web/herschel/pointing-performance}}. The model is then convolved with a Herschel 160 \textmu m point-spread-function (PSF) and re-binned to the 3.2" pixel scale of the Herschel image. A flat background offset is added to the model image before subtraction from the observational data for calculation of the $\chi^2$. Two PSFs are tested, an empirical PSF is adapted from a 160 \textmu m calibration observation of the point source $\gamma$\,Dra from the same Observing Day (Observation IDs 1342231899/1342231900) and a high resolution synthesised PSF from observations of Vesta and Mars provided by \citet{Bocchio2016}. Aside from the $x$ (RA) and $y$ (Dec) image offsets the only other parameters allowed to vary are the radius of the disc, the flux of the disc, the flux of the compact source and the flat background flux of the model. The implementation of a flat background is justified as the annulus of width $\sim$10" (3 pixels) around the detection of Fomalhaut\,C's disc has a median pixel value of $\sim$0.2\,mJy, which the model will need to be able to account for. The disc and compact sources fluxes are allowed to vary as their relative proportions are unknown at the wavelength of observation due to their unknown spectral slopes. The radius of the disc is allowed to vary in order to investigate whether radiation forces and stellar winds from the host star are significant enough to blow out the smaller grains probed by Herschel to larger radii, to probe the potential presence of a small grain halo as \citet{Matthews2015} found for AU\,Mic. We use \textit{emcee} to fit model discs and compact sources to the Herschel data. We use 100 walkers and as we find the largest auto-correlation time across all parameters to be 160 steps, we use 2000 steps and discard the first 1600 steps.

\begin{table}
\centering
\renewcommand{\arraystretch}{1.5}
\caption{Median parameters for the Herschel model with Empirical and Synthesised PSFs. Telescope offsets are measured between the Herschel image coordinate Gaia DR2 stellar location and the model stellar location.}
\label{tab:HerschelResults}
\begin{tabular}{lcccc}
\hline
Parameter & Empirical PSF & Synthesised PSF \\
\hline                                     
Telescope RA Offset (")       & $0.5^{+0.6}_{-0.6}$    & $0.6^{+0.6}_{-0.6}$\\
Telescope Dec Offset (")      & $0.3^{+1.1}_{-1.1}$    & $0.3^{+1.1}_{-1.1}$ \\
Disc Radius (")               & $3.7^{+2.0}_{-2.2}$    & $3.5^{+2.3}_{-2.2}$  \\
Disc Flux (mJy)               & $18^{+7}_{-7}$      & $15^{+6}_{-7}$     \\
Compact Source Flux (mJy)       & $0.9^{+0.6}_{-0.5}$    & $0.9^{+0.5}_{-0.5}$    \\
Background (mJy/arcsec$^{-2}$)         & $0.050^{+0.006}_{-0.006}$    & $0.053^{+0.006}_{-0.006}$     \\
   
\hline
\end{tabular}

\end{table}

The results are summarised in Table\,\ref{tab:HerschelResults} and are
highly consistent between the two PSFs. We find that a small pointing offset is favoured, but within the 1$\sigma $ absolute pointing uncertainty of 1.12". The radius of the disc is not found to be well constrained, but are consistent with the resolved ALMA value. Smaller radii still fit the data well, implying the disc is either unresolved or not substantially resolved with Herschel; radii larger than $\sim5 - 6 $" do not fit the data well. Thus we conclude that there is not sufficient evidence to suggest that the grains probed by Herschel lie at significantly larger radii than the grains probed by ALMA. A flat background flux of  $\sim$0.05\,mJy/arcsec$^{-2}$ is fit by the model, but is not interpreted as significant evidence of a halo of small grains as large amounts of the Herschel map not associated with Fomalhaut\,C share this non-zero flux. The flux fitted to the disc is $\sim$16\,mJy, consistent with the original reported detection of 15.5\,$\pm$\,2.8\,mJy. The compact source is found to only contribute $\sim$0.86\,mJy. We therefore conclude that the Herschel detection of the Fomalhaut\,C disc is not significantly contaminated by the compact source identified by ALMA.

The model flux found for the south-west compact source has a wide uncertainty, and is subject to further systematic uncertainties. But with rough flux estimates at both 160 \textmu m and 870 \textmu m we can estimate a flux ratio of these two wavelengths of $1.5-7.5$. This ratio range would be inconsistent with a sub-millimetre galaxy in the rest frame, but would be consistent with a galaxy at a redshift z = $2-4$ \citep{Casey14}. We note that this redshift range is outside of the sample of z = $1.6\pm 0.4$ used by \citet{Aravena16}, thus causing a potential conflict with the probability estimate that the source is a galaxy. However, we again highlight the systematic uncertainties in he flux derived from the Herschel modelling, if the background source contributed significantly less to the Herschel flux, such a large redshift would not be needed.

\begin{figure}
        {\includegraphics[width=\columnwidth]
        {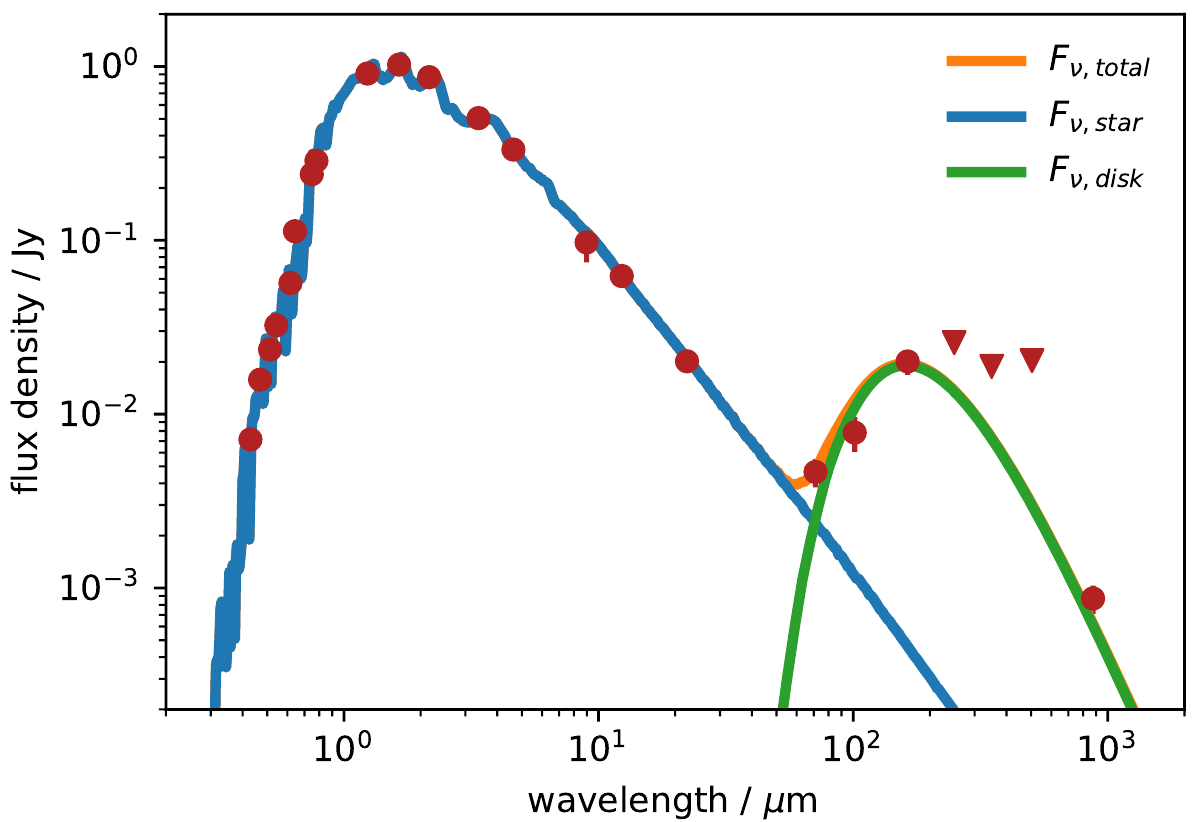}}
        \caption{{\label{fig:sed}}Fomalhaut\,C flux density distribution (SED). Dots are measured fluxes and triangles are 3$\sigma$ upper limits \citep{Kennedy13}. The stellar photosphere model is in blue, the disc model in green and the combined SED in orange.}
\end{figure}

As we have found that the original Herschel flux measurements are consistent with the ALMA findings and that the compact source did not significantly contaminate the detection, those values are kept the same for the fitting of a new blackbody dust model with inclusion of the ALMA flux measurement (see \citet{Yelverton19,Yelverton20} for details of the SED fitting method). The dust model is a modified blackbody spectrum: beyond the fitted parameter $\lambda_0$ there is an additional multiplication factor of $(\lambda/\lambda_0)^{-\beta}$ as small grains do not efficiently radiate at wavelengths larger than their own size. The flux density distribution (SED), Figure\,\ref{fig:sed}, has not been significantly adjusted from \citet{Kennedy13} and the parameters remain consistent. A dust temperature of $20$\,$\pm$\,4\,K is found, corresponding to a blackbody radius (the distance between the dust and the star if the dust grains were perfect blackbodies) of 13\,$\pm$\,5\,au, with a fractional luminosity of $L_{\rm dust}/L_{\rm \star} = 1.5 \pm 0.2 \times 10^{-4}$. While $\lambda_0$ is not well constrained, we find $\beta$ = $1.5 \pm 0.4$. 

\subsection{Blackbody vs Resolved Radii}\label{sec:gamma}

With the newly resolved radius of the disc of 26\,au the blackbody radius of 13\,au can be seen to be a significant underestimate. This is a common finding for debris discs around all host stellar types \citep{Rodriguez12,Booth13,Pawellek15} implying the presence of small dust grains that are hotter than black bodies due to their inefficient long wavelength emission. We can use a measure of this called $\Gamma$, defined as R$_{\rm dust}/$R$_{\rm BB}$, the ratio of the resolved disc to the blackbody radius \citep{Booth13}, or equivalently defined as $\rm (T_{\rm dust}/\rm T_{\rm BB})^2$, the square of the ratio of the dust temperature to the temperature of an ideal blackbody at the radius of the disc \citep{Pawellek15}. The $\Gamma$ factor for Fomalhaut\,C's disc is 1.9\,$\pm$\,0.7.

The loose trend \citep{Pawellek15} is that $\Gamma$ increases with decreasing stellar luminosity, albeit with strong scatter. 
This trend is linked to typical grain sizes decreasing towards stars with lower luminosities exhibiting weaker radiation pressure on the dust, i.e. the blowout size and with it the minimum grain size $D_{\rm min}$ decreases with decreasing stellar luminosity.
Smaller grains are typically hotter than larger grains due to decreased emission efficiency and thus the blackbody discrepancy grows. Working from \citet{Pawellek15}'s relations Fomalhaut\,C's ($L_{\text{Fom\,C}}\approx0.005\,L_{\sun}$) $\Gamma$ should be larger at $\sim5-12$, more similar to that measured for GJ\,581 ($L_{\text{GJ\,581}}\approx0.01\,L_{\sun}$). As it stands Fomalhaut\,C's $\Gamma$ is even smaller than the $\sim3-4$ of AU\,Mic ($L_{\text{AU\,Mic}}\approx0.1\,L_{\sun}$). However there are a couple of key caveats aside from the large observed scatter. 

Firstly, visible light absorption efficiency significantly decreases for smaller astrosilicate particles \citep[$\lesssim$\,microns;][]{Krivov08} serving to plateau the trend of increasing dust temperature with decreasing grain size as decreasing absorption efficiency begins to counter the decreasing emission efficiency. Around lower temperature stars this turnover would be reached at comparatively larger minimum grain sizes as the peak stellar emission is moved to longer wavelengths. 
Secondly, as radiation pressure from low mass stars begins to become too weak to remove grains altogether, Poynting-Robertson drag (P-R drag) and stellar wind \citep[e.g.][]{Wyatt11,Reidemeister11,Plavchan05} become the dominant grain-removal mechanisms. The radiation pressure dominated trend of decreasing $D_{\rm min}$ with decreasing stellar luminosity is now disrupted and it is unclear how the relationship proceeds to lower luminosities.
As a very low luminosity star these effects would be particularly prominent for Fomalhaut\,C and could explain why \citet{Pawellek15}'s $\Gamma$ trend has appeared to have flattened or possibly even turned over in the low mass regime in which Fomalhaut\,C belongs. Aside from Fomalhaut\,C, AU\,Mic is the only other M\,star currently thermally resolved in high resolution \citep[TWA\,7 is only marginally resolved with ALMA and GJ\,581 is only marginally resolved with Herschel;][respectively]{Bayo19,Lestrade12}, as more are resolved with ALMA it will be valuable to investigate $\Gamma$ and grain sizes in this low mass regime of low temperature hosts, stellar wind and small grains.

\subsection{Scattered Light Non-Detections}\label{sec:scattered}
\subsubsection{HST/STIS Observations}

\begin{figure*}
        \centering
        \includegraphics[width=0.325\linewidth]{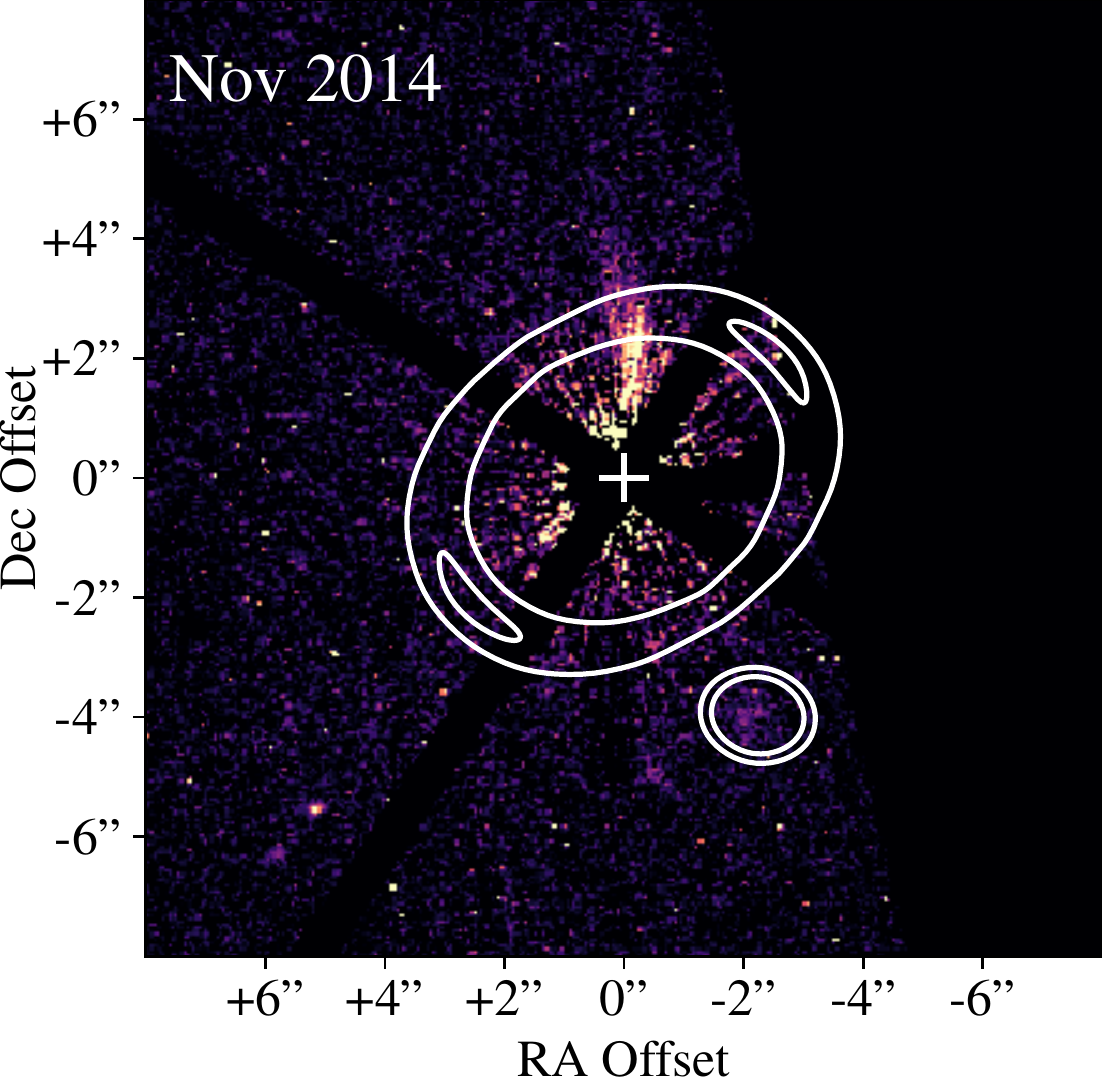}
        \hfill
        \includegraphics[width=0.325\linewidth]{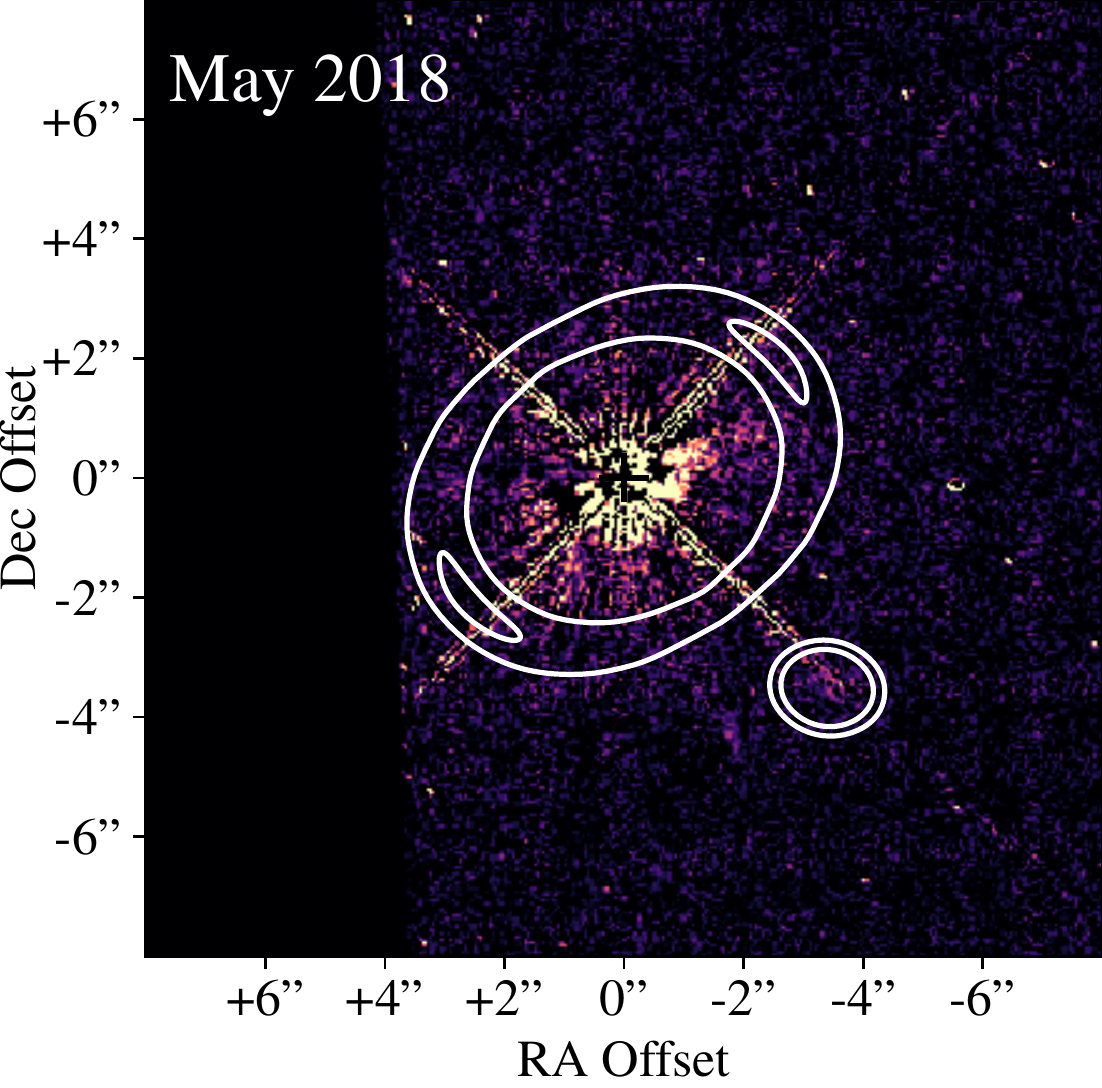}
        \hfill
        \includegraphics[width=0.325\linewidth]{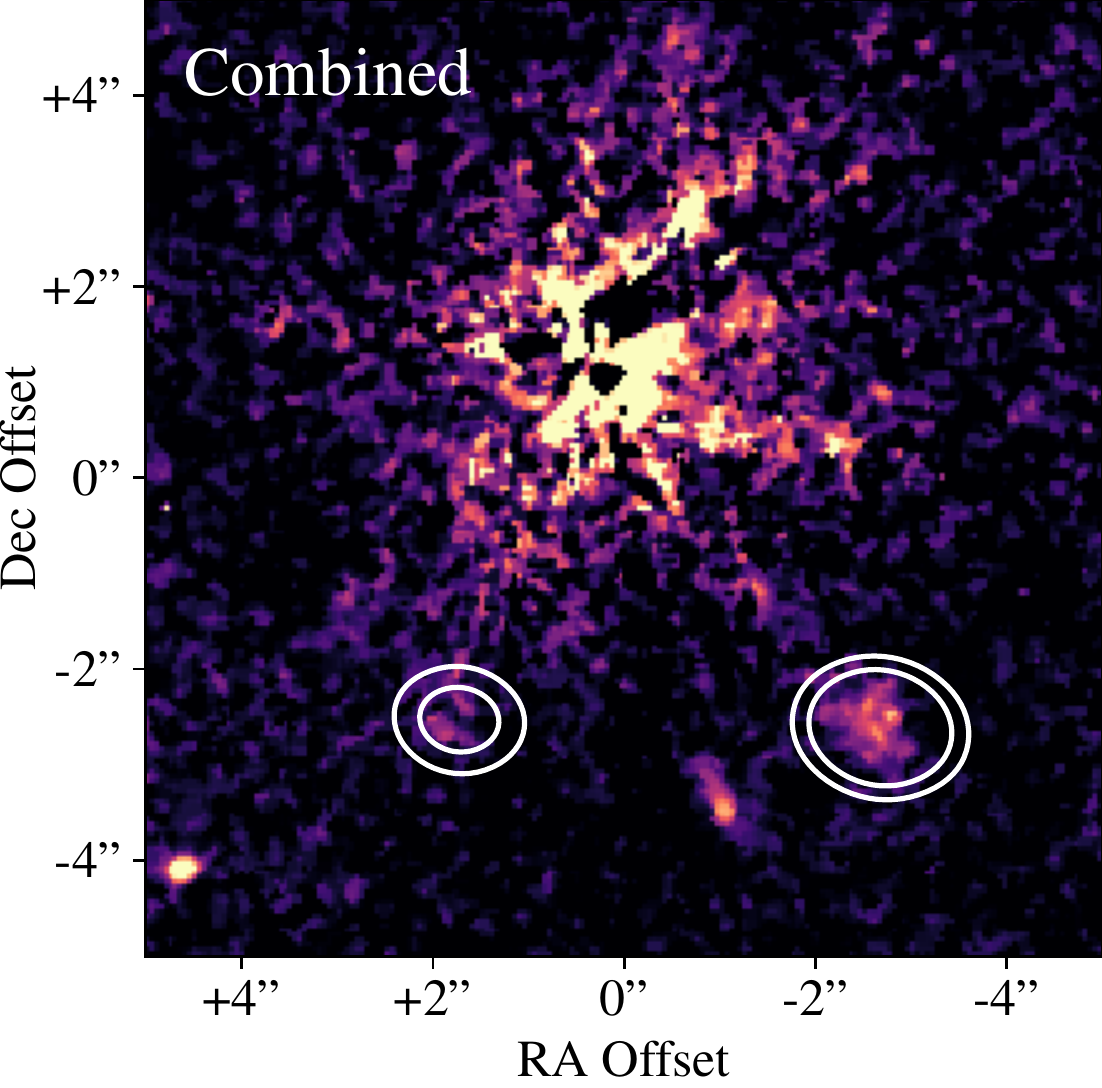}
    \caption{\label{fig:Hubble}HST STIS observations of Fomalhaut\,C. Left: Initial November 2014 observation with diffuse linear structure apparent north of the star. The disc is not detected but the background source found by ALMA external to the disc in the south-west is also detected by STIS. In white are contours of $1, 2,\sigma$ from the ALMA best fit torus model, assuming no proper motion of the external compact source. The Gaia DR2 location of the star at the epoch of observation is marked with a $+$. Zero offset is at 342$^\circ$01'12.8" $-$24$^\circ$22'10.5" (J2000). Middle: May 2018 observation with diffuse linear structure north of the star not detected. The disc is not detected and the background source external to the disc is potentially obscured by the telescope's diffraction spikes. Overlays as before. Zero offset is at 342$^\circ$01'14.1" $-$24$^\circ$22'11.1" (J2000). Right: Combined observations of all HST epochs to increase background object SNR, Fomalhaut\,C is blurred due to high proper motion and thus we do not plot the disc model contours. The exterior compact source is detected in the south-west, but a potential in-disc point source in the south-east is not detected. In white are contours of $1, 2,\sigma$ of the two compact sources, within the disc and exterior to the disc, from the ALMA best fit torus with point source model assuming no proper motion. Zero offset is at 342$^\circ$01'13.0" $-$24$^\circ$22'11.6" (J2000). }
\end{figure*}

We attempted to detect dust-scattered light around Fomalhaut\,C using HST/STIS coronagraphy. STIS comprises a 1024x1024 pixel CCD with various occulting elements in the focal plane and a scale of 0.05077 "/pixel. However, STIS does not have a pupil-plane mask to suppress the four diffraction spikes from HST's secondary support structure, nor does it have filters, effectively operating at the wide optical throughput of the system ($\lambda_c = 0.5858~\mu$m, $\Delta \lambda = 0.4410~\mu$m). 

Fomalhaut\,C was observed at two epochs: UT 2014-11-12 (HST-GO-13725) and 2018-05-28 (HST-GO-15172) as shown in Figure \ref{fig:Hubble} left and middle respectively. At each epoch Fomalhaut\,C was occulted behind BAR5 with width $\sim$0.4", and observed at two telescope roll angles separated by $\sim$30$\degr$ in two consecutive orbits. Each orbit in the 2014 epoch comprised six exposures of 397 seconds whereas the 2018 epoch had six 379 second exposures. Cosmic rays were removed in each *flt.fits exposure by interpolation over the bad pixels identified in the *.pl files and then the six exposures per orbit were median combined. The sky background was sampled in a region on the detector farthest away from the occulted star and subtracted. Finally the images were divided by the integration time. To subtract the point-spread function, the final image from the first orbit was iteratively shifted and subtracted from the second orbit.

The 2014 data revealed a diffuse, nearly-linear structure extending northward from Fomalhaut\,C between 1.2" (the edge of the occulted region) and 3" radius. The morphology and surface brightness resembled a background galaxy also seen 18.4" to the east of Fomalhaut\,C, highlighting the possibility that the Fomalhaut\,C extended feature was also a background galaxy.This finding motivated the 2018 observations in order to check for common proper motion of the feature with the star. However, the feature was not detected anywhere in the 2018 field, showing it to be a spurious artifact in the 2014 data.

No circumstellar nebulosity is detected in the STIS data with a 3$\sigma$ limited surface brightness of 3.39\,$\mu$Jy arcseconds$^{-2}$ at 3" radius from the star (using a zero point of 1 DN/s/pixel = 4.55x10$^{-7}$\,Jy).
The 7$\sigma$ ALMA compact source south-west of Fomalhaut\,C is detected in the 2014 observation (Figure \ref{fig:Hubble} left). It may also be detected in the 2018 observation (Figure \ref{fig:Hubble} middle), but its location is obscured by the telescope's diffraction spikes. The ALMA source is significantly detected in a combined image of all HST observations (Figure \ref{fig:Hubble} right), showing that it is indeed real. These detections together demonstrate that the source does not share Fomalhaut\,C's proper motion and is a background object. In neither epoch nor in the combined image is there a significant detection of a potential south-east point source within the disc per the ALMA torus with point source model.

\subsubsection{VLT/SPHERE Observations}

\begin{figure*}
        {\includegraphics[width=2\columnwidth]
        {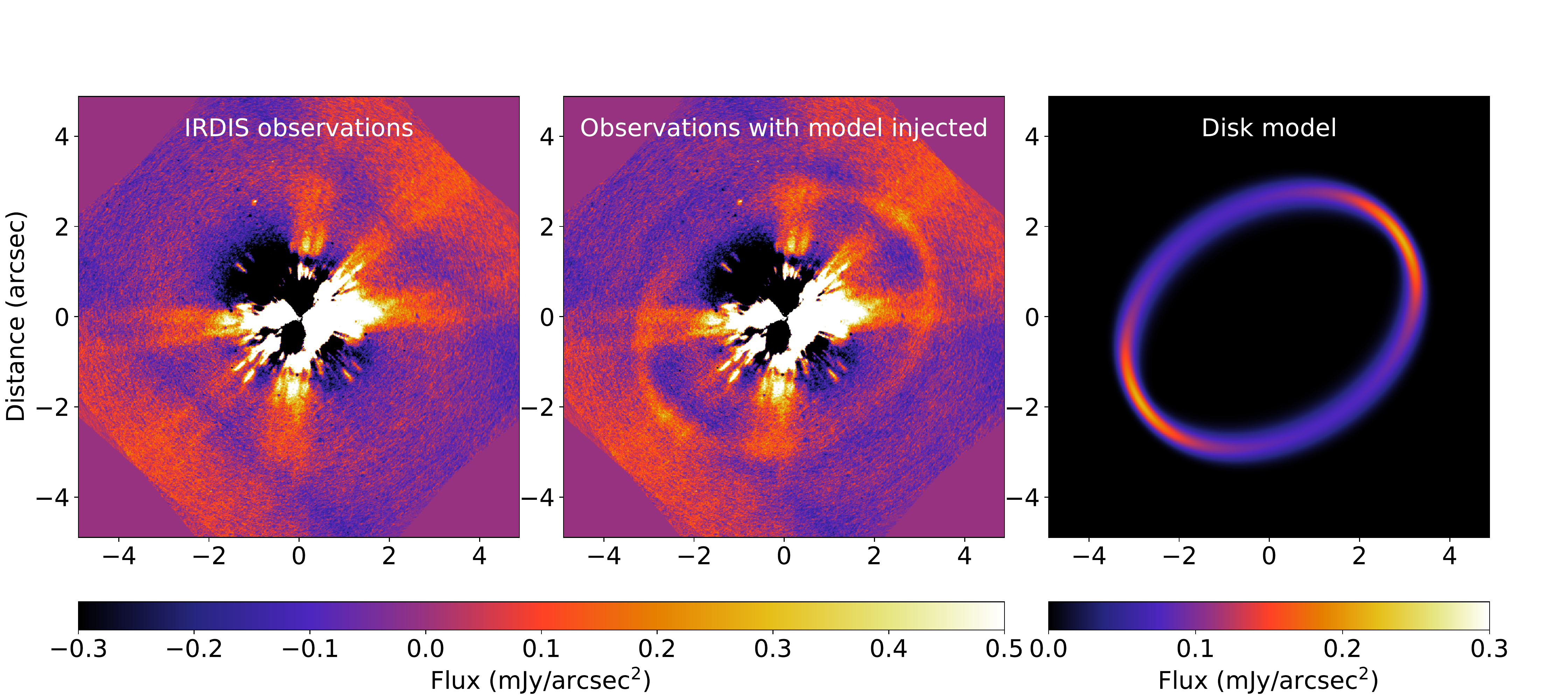} }
        \caption{\label{fig:FomC_IRDIS} Left: final reduced image after combinations of two epochs of observation with SPHERE/IRDIS. The radially extended signal seen at four different position angles originates from the diffraction pattern of the telescope spiders, which are not entirely suppressed by the pupil stop of the Lyot coronagraph. Middle: final reduced image after injection of a disc model into the data before post-processing. Right: scattered light model injected in the data.}
\end{figure*}

Fomalhaut\,C was observed with the high-contrast imager VLT/SPHERE \citep{Beuzit2019} as part of the SPHERE High-Angular Resolution Debris Disks Survey\footnote{ESO programs 096.C-0388 and 097.C-0394} \citep[SHARDDS,][]{Wahhaj2016,Choquet2017,Marshall2018}. This survey is an imaging search for discs around stars within 100 pc having an infrared excess greater than $10^{-4}$. It uses the IRDIS subsystem \citep{Dohlen2008} in broad band H ($\lambda=1.625\,\mu$m, $\Delta\lambda=0.290\,\mu$m) and the apodised Lyot coronagraph of diameter 185 mas. Fomalhaut\,C was observed at 2 epochs on the nights of 11 October 2015 and 3 June 2016, with an exposure time of 40 minutes on-source for each visit. The observations were carried out in pupil-stabilised mode, however very little sky rotation was obtained (only $1.5^\circ$) because the target was observed outside the meridian crossing. Angular Differential Imaging \citep[ADI,][]{Marois2006} is therefore not practical due to severe self-subtraction of any astrophysical signal \citep[e.g.][]{Milli2012}. At the expected separation of the disc of $\sim$3.4", the background noise is the main contributor to the noise. We therefore derotated the frames to align North vertically on the detector, subtracted the median azimuthal profile for each frame at each separation, and median-combined all frames to obtain the final reduced images at each epoch. We averaged the reduced images of the two epochs to produce the final image shown in Figure \ref{fig:FomC_IRDIS} (left). The disc is not detected in scattered light in the IRDIS image and we calculate a $3\sigma$ surface brightness detection limit of 173\,$\mu$Jy arcsecond$^{-2}$ at 3.4" from the star, assuming a stellar brightness in the H band of 7.527 mag \citep[1.01 Jy; 2MASS][]{Cutri2003}. Post processing could change this value, so to derive meaningful constraints based on the ALMA detection, we used the median parameter Gaussian torus model of the disc to generate an image (Figure\,\ref{fig:FomC_IRDIS} right) that we injected into the SPHERE data. We then scaled up the image until it was clearly detected in the final reduced image (Figure\,\ref{fig:FomC_IRDIS} middle). We find that the integrated brightness of the model is 1.0\,mJy, which represents an upper limit on the total disc scattered light brightness. The maximum surface brightness reached at the ansae of the disc is $\sim$ 200\,$\mu$Jy arcsec$^{-2}$, showing that our $3\sigma$ surface brightness detection limit is reasonable. 

\subsubsection{Limits on dust albedo}
 That the disc was not detected with either HST/STIS or VLT/SPHERE could be owing to its unfavourable viewing geometry. At an inclination of $\sim 43^\circ$ low scattering angles are unavailable to observation, which leaves the strong forward scattering peak that can enhance disc surface brightness inaccessible. However, an upper limit on the dust albedo can still be calculated from the surface brightness upper limit. Following the process outlined in section 3.3.3 of \citet{Marshall2018}, we use the equation for reflection for optically thin dust from \citet{weinberger1999}: 
 \begin{equation}
 \label{eqn:scattered 1}
      \tau\omega = 4\pi\phi^2 \frac{S}{F}
 \end{equation}
where $\tau$ is the optical depth, $\omega$ is the albedo, $\phi$ is the angular separation of the scatterers from the host star (i.e. the disc semi-major axis in arcseconds), $S$ is the surface brightness of the disc in mJy\,arcsecond$^{-2}$ and $F$ the stellar flux in mJy. We also use the approximation for optical depth:
 \begin{equation}
 \label{eqn:scattered 2}
      \tau = \frac{2f\phi \cos(i)}{d\phi(1-\omega)}
 \end{equation}
where $f$ is the fractional luminosity of the disc, $i$ the inclination and $d\phi$ the disc width in arcseconds. We combine the two to eliminate $\tau$ and have:
 \begin{equation}
 \label{eqn:scattered 3}
      S = \frac{fF\omega}{(2\pi\phi d\phi \cos(i))(1-\omega)}
 \end{equation}
 into which we can insert our Gaussian torus model values from \S\ref{sec:Gaussian Torus} and surface brightness upper limits to extract our albedo upper limit. From the SPHERE observations we retrieve an albedo upper limit of 0.67 at 1.625\,$\mu$m and from the STIS observations we retrieve an upper limit of 0.54 at 0.5858\,$\mu$m. Typical debris disc dust albedos range between 0.05--0.15 \citep[e.g.][]{Marshall2018,Choquet18,Golimowski11,Krist10,kalas2005} and typical Kuiper belt objects have average albedos of 0.11 -- 0.17 \citep{Vilenius2012}; precision on this level is needed to begin distinguishing between compositional models \citep{Marshall2018,Choquet18}. Thus these upper limits are too high to comment on dust composition.
 
\section{Discussion}
\label{sec:discussion}
In our ALMA observations we do not find evidence for a significant eccentricity in Fomalhaut\,C's disc, and it is most likely that Fomalhaut\,C has a less eccentric disc than Fomalhaut\,A. In the context of S14's models, the system's history thus remains inconclusive.
Figure\,6 in S14 shows that in their scenario the eccentricity of Fomalhaut\,A's disc should be correlated with the eccentricity of C's disc, but with a large scatter. No definite prediction could be made for an eccentricity in C's disc; in S14's Figure\,6 it can be seen that for A discs reminiscent of the real A with eccentricities between 0.025 and 0.5, the corresponding C disc eccentricities vary between $\sim $0.025 -- 0.75.
Thus, our 3$\sigma$ upper limit of 0.14 cannot rule out the S14 history. However, if an eccentricity did exist below this limit, at such a magnitude the origin of the eccentricity could just as much be attributed to other factors, such as an eccentric planet within the system. A larger eccentricity would have been more unusual, thus implying an unusual cause, i.e. the S14 scenario. Further observation increasing the S/N and the precision of the offset modelling would not therefore necessarily help break the degeneracy of the Fomalhaut system's potential histories,
however deeper observations can also reveal other observable quantities such as the dust density distribution that can also trace system dynamics.

The still indefinite history of the Fomalhaut system precludes a ruling on the 'typicality' of the Fomalhaut\,C debris disc's brightness amongst the M\,star disc population, as the disc could have been additionally stirred by gravitational interactions with Fomalhaut\,A per S14's scenario.
With a fractional luminosity of 1.5$\times 10^{-4}$ the disc is certainly very bright, on par with the disc around AU\,Mic, an earlier type M0 star that is $\sim$1/40 times the age. AU\,Mic will not retain its current brightness for the next 400 Myr, as disc mass tends to decrease over time due to collisional grinding of planetesimals and removal of dust from the system through radiation pressure, solar winds and P-R drag \citep{Wyatt08}. This does not mean that Fomalhaut\,C's disc was necessarily significantly brighter in the past as the time of onset of its collisional cascade is unknown. It is still within the realm of possibility that the Fomalhaut\,C debris disc originated from a protoplanetary disc that formed a greater mass in planetesimals than AU\,Mic's and evolved to its current state via natural collisional grinding. While Fomalhaut\,C is significantly older than AU\,Mic, in comparison to field M\,stars as a whole Fomalhaut\,C is still young at only 440 Myr old compared to ages ranging up to 10 Gyrs. A proper study of disc occurrence and comparison for M\,stars could select a sample of stars of similar ages, preferably young while the discs are statistically likely to be brightest.

\begin{figure}
        {\includegraphics[width=\columnwidth]
        {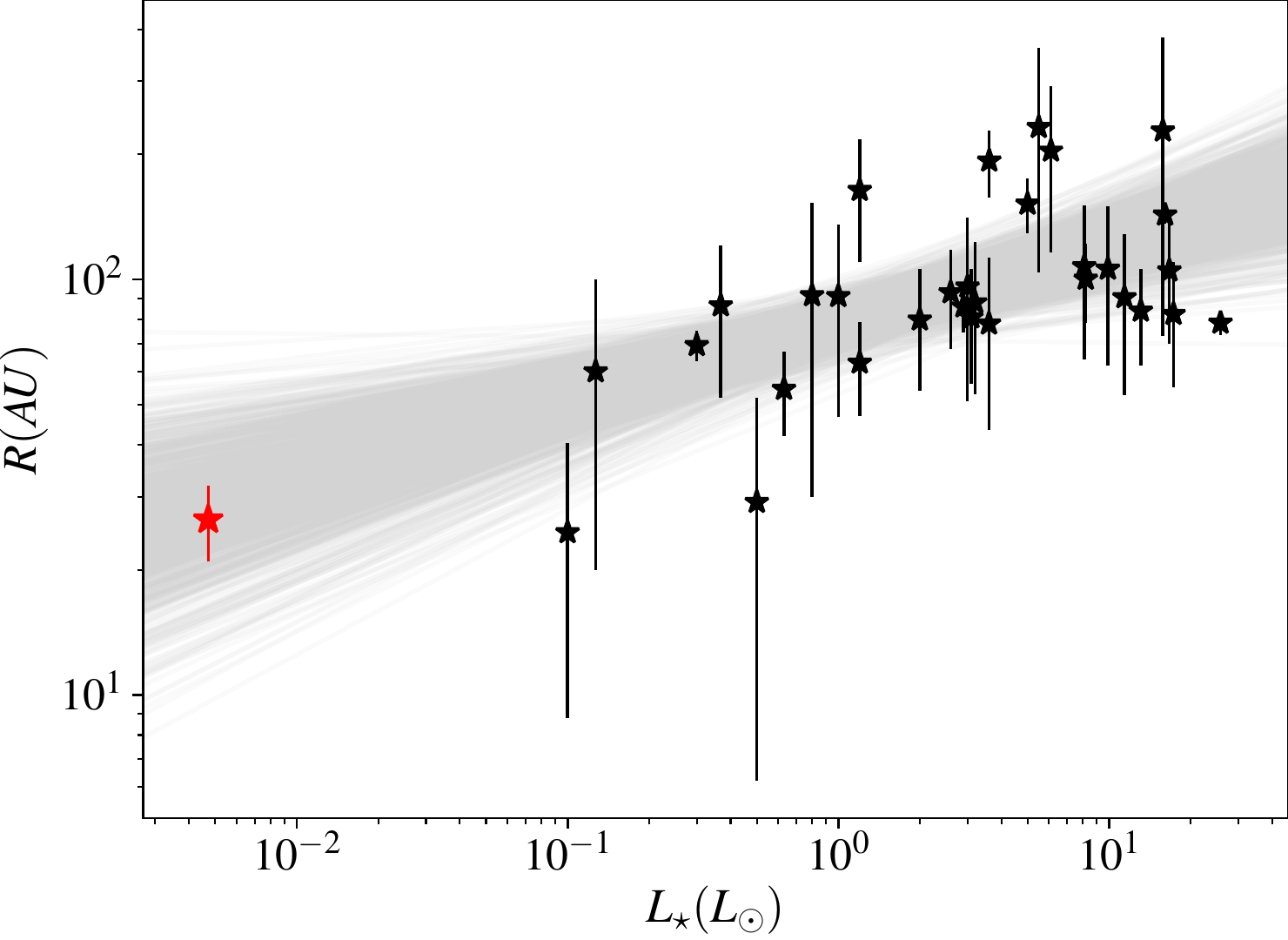}}
        \caption{\label{fig:LRPlot} Resolved planetesimal belt radii against stellar host luminosity. The error bars represent disc widths or the upper limits thereof. Fomalhaut\,C is highlighted in red, the error bar represents the 3$\sigma$ limit on the disc's FWHM. Grey lines show 1000 power laws sampled from the parameter distributions of \citet{Matra2018}.}
\end{figure}

It is however possible and useful to compare the Fomalhaut\,C disc's radius with that of other resolved debris discs. \citet{Matra2018} find a correlation between disc radius and host star luminosity, but their sample is truncated at the low luminosity end, having no discs with hosts of lower luminosity than AU\,Mic, at 0.1\,$L_\odot$. Fomalhaut\,C has a luminosity of 0.005\,$L_\odot$ and thus significantly extends the range of the parameter space. In Figure\,\ref{fig:LRPlot} we plot \citet{Matra2018}'s sample, with the updates from \citet{Sepulveda19} and references therein, and with the addition of Fomalhaut\,C. We also plot a representative sample of power law fits from the parameter distributions calculated by \citet{Matra2018}. The radius of the disc around Fomalhaut\,C is found to be wholly consistent with the rest of the sample, lying close to the centre of the bundle of representative power laws. At least in relation to radius, the Fomalhaut\,C disc appears typical, but more discs around low luminosity host stars are required to fill out this region of the parameter space in order to be able to conclude that the relationship holds.

Regarding Figure\,\ref{fig:LRPlot} we are reminded of the context of the width of the Fomalhaut\,C disc; the disc is relatively narrow, similar to the likes of $\epsilon$\,Eri \citep{Booth17}, HR\,4796A \citep{Kennedy18b} and indeed Fomalhaut\,A \citep{kalas2005,Acke2012,macgregor17}. Narrow rings are very often also offset from the stellar location \citep{Hughes18}. The typical eccentricities of these narrow discs are $\sim$0.1 and so the non-detection of an eccentricity in Fomalhaut\,C's disc not does not mark it as unusual for its narrowness. Postulated reasons for narrow rings can be similar to those for eccentric discs: shepherding planets for the inner and outer radii \citep{Boley12}, this would predict sharp edges that our current resolution is unable to constrain; confinement by the orbital resonances of a single planet, like the bounds of the Kuiper belt, another narrow disc, at the 3:2 and 2:1 resonances with Neptune \citep{Hahn05}; or dust-gas interaction mechanisms \citep{Lyra13}. If the bounds of Fomalhaut\,C's disc correspond to 3:2 and 2:1 resonances of an unseen planet, our model would suggest a planet at an orbital distance of $\sim$17--20 au.

\section{Conclusions}
In this work we have presented the first resolved sub-millimetre observations of the planetesimal debris disc around Fomalhaut\,C (\S\ref{sec:ICA}), now the lowest mass star to have its disc resolved in thermal emission. Our modelling has revealed the geometry of the ring as well as its radius and submillimetre flux. We try three distinct models to investigate the nature of the over-brightness in the south-east quadrant of the disc and conclude that the symmetric Gaussian torus model is the best fitting (\S\ref{sec:ModSum}). We search for an offset of the centre of the disc from the stellar location but do not find any significant eccentricity, instead placing a 3$\sigma$ upper limit. Higher signal-to-noise and/or resolution observations will be necessary to improve the precision of an offset measurement and to measure the disc's scale width and height. We also do not detect any CO gas in the system but place a 3$\sigma$ upper limit of 17 mJy\,km\,s$^{-1}$.

We revisit the original Herschel observations with our best-fitting ALMA model to consider a scenario where the smaller grains visible at shorter wavelengths lay at larger radii due to radiation pressure and stellar wind forces blowing them out, but do not find evidence for a small grain halo. We can conclude however that the original Herschel observations were not significantly contaminated by the compact source apparent outside of the ring in the ALMA observations (\S\ref{sec:Herschel and Revised SED Model}). 

With the ring's radius resolved we compare the disc's blackbody radius to its resolved radius to calculate $\Gamma=$\,R$_{\rm dust}/$R$_{\rm BB}$ and compare it to discs around stars of other spectral types. We find that Fomalhaut\,C's $\Gamma$ factor is smaller than might be expected from the trends of earlier type stars but also outline several caveats that could disrupt the trends for very low mass stars (\S\ref{sec:gamma}).

The Fomalhaut\,C disc has not been detected in scattered light with either HST/STIS in the optical or VLT/SPHERE in the near-infrared, but we use our ALMA model's geometry to find upper limits on surface brightness and dust albedo.
These limits are not constraining enough to investigate different dust composition models (\S\ref{sec:scattered}).

The lack of a significant offset measurement precludes a judgement on the likelihood on any particular dynamical history model for the Fomalhaut triple system. In combination with the paucity of thermally resolved M\,star debris discs this uncertainty in history makes it difficult to rule on the disc's typicality or to place it within the context of low mass star discs. We do place it in the context of debris discs across all types by adding it to \citet{Matra2018}'s radius-luminosity sample and find that the Fomalhaut\,C's disc radius is entirely consistent with the greater trend (\S\ref{sec:discussion}). 

\section*{Acknowledgements}
PFCC is supported by the University of Warwick.
GMK is supported by the Royal Society as a Royal Society University Research Fellow.
GD acknowledges support from NSF AST-1518332 and NASA NNX15AC89G and NNX15AD95G/NExSS. 
CJC acknowledges support the European Union's Horizon 2020 research and innovation programme under the Marie Sklodowska-Curie grant agreement No 823823 (DUSTBUSTERS) and STFC Consolidated Grant ST/S000623/1.
This paper makes use of the following ALMA
data: ADS/JAO.ALMA\#2017.1.00561.S. ALMA is a partnership
of ESO (representing its member states), NSF (USA) and NINS
(Japan), together with NRC (Canada), MOST and ASIAA (Taiwan),
and KASI (Republic of Korea), in cooperation with the Republic
of Chile. The Joint ALMA Observatory is operated by ESO,
AUI/NRAO and NAOJ.
This research makes use of observations with the NASA/ESA
Hubble Space Telescope, which is operated by the Association of Universities
for Research in Astronomy.
This work has made use of data from the European Space Agency (ESA) mission
{\it Gaia} (\url{https://www.cosmos.esa.int/gaia}), processed by the {\it Gaia}
Data Processing and Analysis Consortium (DPAC,
\url{https://www.cosmos.esa.int/web/gaia/dpac/consortium}). Funding for the DPAC
has been provided by national institutions, in particular the institutions          
participating in the {\it Gaia} Multilateral Agreement.
We sadly miss Professor Wayne Holland of the Royal Observatory of Edinburgh who was a leading figure in sub-millimetre astronomy and enthusiastically joined this project from its start. This paper is dedicated to his memory.

\section*{Data Availability}
The data underlying this article
are available in http://almascience.nrao.edu/aq/,
https://archive.stsci.edu/index.html, and  http://archive.eso.org/cms.html and can be accessed with ALMA
project ID: 2017.1.00561.S;
HST proposal IDs: HST-GO-13725 and HST-GO-15172; and ESO program IDs: 096.C-0388 and 097.C-0394.


\bibliographystyle{mnras}
\bibliography{fomcBib}




\bsp	
\label{lastpage}
\end{document}